\let\includefigures=\iftrue
%
\let\useblackboard=\iftrue
%
%
\newfam\black
\input harvmac.tex
\input xy
\xyoption{all}

\includefigures
\message{If you do not have epsf.tex (to include figures),}
\message{change the option at the top of the tex file.}
\input epsf
\def\figin{\epsfcheck\figin}\def\figins{\epsfcheck\figins}
\def\epsfcheck{\ifx\epsfbox\UnDeFiNeD
\message{(NO epsf.tex, FIGURES WILL BE IGNORED)}
\gdef\figin##1{\vskip2in}\gdef\figins##1{\hskip.5in}
\else\message{(FIGURES WILL BE INCLUDED)}%
\gdef\figin##1{##1}\gdef\figins##1{##1}\fi}
\def\DefWarn#1{}
\def\figinsert{\goodbreak\midinsert}
\def\ifig#1#2#3{\DefWarn#1\xdef#1{fig.~\the\figno}
\writedef{#1\leftbracket fig.\noexpand~\the\figno}%
\figinsert\figin{\centerline{#3}}\medskip\centerline{\vbox{\baselineskip12pt
\advance\hsize by -1truein\noindent\footnotefont{\bf Fig.~\the\figno:} #2}}
\bigskip\endinsert\global\advance\figno by1}
\else
\def\ifig#1#2#3{\xdef#1{fig.~\the\figno}
\writedef{#1\leftbracket fig.\noexpand~\the\figno}%
\global\advance\figno by1}
\fi
\useblackboard
\message{If you do not have msbm (blackboard bold) fonts,}
\message{change the option at the top of the tex file.}
\font\blackboard=msbm10 scaled \magstep1
\font\blackboards=msbm7
\font\blackboardss=msbm5
\textfont\black=\blackboard
\scriptfont\black=\blackboards
\scriptscriptfont\black=\blackboardss
\def\Bbb#1{{\fam\black\relax#1}}
\else
\def\Bbb{\bf}
\fi
%
\def\yboxit#1#2{\vbox{\hrule height #1 \hbox{\vrule width #1
\vbox{#2}\vrule width #1 }\hrule height #1 }}
\def\fillbox#1{\hbox to #1{\vbox to #1{\vfil}\hfil}}
\def\ybox{{\lower 1.3pt \yboxit{0.4pt}{\fillbox{8pt}}\hskip-0.2pt}}
\def\subsubsec#1{\medskip
\noindent {\it #1}
\medskip}
%

%

\def\bbbone{{\mathchoice {\rm 1\mskip-4mu l} {\rm 1\mskip-4mu l}
          {\rm 1\mskip-4.5mu l} {\rm 1\mskip-5mu l}}}

\def\comments#1{}

\def\QH{\Bbb{H}}

\def\p{\partial}

\def\Tr{{{\rm Tr~ }}}

\def\Re{{\rm Re\hskip0.1em}}
\def\Im{{\rm Im\hskip0.1em}}

\def\cF{{\cal F}}

\def\cM{{\cal M}}

\def\cV{{\cal V}}

\def\II{\relax{I\kern-.10em I}}

\def\IIb{{\II}b}

\def\IZ{\relax\ifmmode\mathchoice
{\hbox{\cmss Z\kern-.4em Z}}{\hbox{\cmss Z\kern-.4em Z}}
{\lower.9pt\hbox{\cmsss Z\kern-.4em Z}}
{\lower1.2pt\hbox{\cmsss Z\kern-.4em Z}}\else{\cmss Z\kern-.4em
Z}\fi}
\def\IB{\relax{\rm I\kern-.18em B}}
\def\IC{{\relax\hbox{$\inbar\kern-.3em{\rm C}$}}}
\def\ID{\relax{\rm I\kern-.18em D}}
\def\IE{\relax{\rm I\kern-.18em E}}
\def\IF{\relax{\rm I\kern-.18em F}}
\def\IG{\relax\hbox{$\inbar\kern-.3em{\rm G}$}}
\def\IGa{\relax\hbox{${\rm I}\kern-.18em\Gamma$}}
\def\IH{\relax{\rm I\kern-.18em H}}
\def\II{\relax{\rm I\kern-.18em I}}
\def\IK{\relax{\rm I\kern-.18em K}}
\def\IN{\relax{\rm I\kern-.18em N}}
\def\IP{\relax{\rm I\kern-.18em P}}

%
\def\inbar{\,\vrule height1.5ex width.4pt depth0pt}

\def\p{\partial}

\font\cmss=cmss10 \font\cmsss=cmss10 at 7pt
\def\IR{\relax{\rm I\kern-.18em R}}

\def\CM{{\cal M}}
\def\CN{{\cal N}}

\def\BR{\IR}
\def\BZ{\IZ} 
\def\BP{\IP}
\def\BR{\IR}
\def\BC{\IC}

\def\BH{\QH}

\def\lp10{l_P^{10}}
\def\lp11{l_P^{11}}
\def\R11{R_{11}}
\def\tV{\tilde{V}}
\def\tD{\tilde{D}}

\Title{\vbox{\baselineskip12pt\hbox{hep-th/0404257}}}
{\vbox{\vskip 37pt
\centerline{Building a Better Racetrack}}}
\smallskip
\centerline{Frederik Denef, Michael R. Douglas\footnote{$^{\&}$}{
Louis Michel Professor}$^{,2}$ and Bogdan Florea}
\medskip
\medskip
\centerline{{NHETC and Department of Physics and Astronomy,}}
\centerline{{Rutgers University, Piscataway, NJ 08855--0849, USA}}
\medskip
\centerline{{$^\&$I.H.E.S., Le Bois-Marie, Bures-sur-Yvette, 91440 France}}
\medskip
\centerline{{$^2$California Institute of Technology, Pasadena CA 91125, USA}}
\medskip
\centerline{\tt denef, mrd, florea@physics.rutgers.edu}
\bigskip
\bigskip
\bigskip
\noindent
We find \IIb\
compactifications on Calabi-Yau orientifolds in which all K\"ahler
moduli are stabilized, along lines
suggested by Kachru, Kallosh, Linde and Trivedi.

\Date{April 2004}
\def\np{{\it Nucl. Phys.}}

\def\pr{{\it Phys. Rev.}}
\def\pl{{\it Phys. Lett.}}
\def\atmp{{\it Adv. Theor. Math. Phys.}}
\def\cqg{{\it Class. Quant. Grav.}}
\def\cmp{{\it Comm. Math. Phys.}}
\nref\achflux{B. Acharya, ``A Moduli Fixing Mechanism in M theory,'' [arXiv:hep-
th/0212294].}

\nref\ad{ S.~Ashok and M.~R.~Douglas, ``Counting Flux Vacua,''
JHEP {\bf 0401}, 060 (2004) [arXiv:hep-th/0307049].}

\nref\aspdoug{P. S. Aspinwall and M. R. Douglas, ``D-Brane Stability and
Monodromy'', JHEP {\bf 0205}, 031 (2002)
[arXiv:hep-th/0110071].}

\nref\bala{V. Balasubramanian,
``Accelerating Universes and String Theory,''\cqg\ {\bf 21}, 1337 (2004)
[arXiv:hep-th/0404075].}

\nref\BanksES{
T.~Banks, M.~Dine and E.~Gorbatov, ``Is there a String Theory Landscape?,''
[arXiv:hep-th/0309170].
}

\nref\banks{T.~Banks,
``Cosmological Breaking of Supersymmetry or Little Lambda Goes Back to the
Future. II,''
[arXiv:hep-th/0007146].}

\nref\BanksES{
T.~Banks, M.~Dine and E.~Gorbatov,
``Is there a string theory landscape?,''
[arXiv:hep-th/0309170].}

\nref\barrow{J. D. Barrow, {\it The Book of Nothing,}
London: Vintage (2001)}

\nref\B{V. Batyrev, ``Dual Polyhedra and Mirror Symmetry for Calabi-Yau
Hypersurfaces
in Toric Varieties,'' {\it J. Alg. Geom.} {\bf 3}, 493 (1994).}

\nref\beckerG{K. Becker and M. Becker, ``M-Theory on Eight-Manifolds,'' \np\ B
{\bf 477}, 155 (1996)
[arXiv:hep-th/9605053].}

\nref\BeckerNN{K.~Becker, M.~Becker, M.~Haack and J.~Louis,
``Supersymmetry Breaking and $\alpha'$-Corrections to Flux Induced
Potentials,'' JHEP {\bf 0206}, 060 (2002) [arXiv:hep-th/0204254].}

\nref\BinetruyHH{
P.~Binetruy, G.~Dvali, R.~Kallosh and A.~Van Proeyen,
``Fayet-Iliopoulos Terms in Supergravity and Cosmology,''
[arXiv:hep-th/0402046].
}

\nref\boupol{R.~Bousso and J.~Polchinski, ``Quantization of
Four-Form Fluxes and Dynamical Neutralization of the  Cosmological
Constant,'' JHEP {\bf 0006}, 006 (2000) [arXiv:hep-th/0004134].}

\nref\breit{P.~Breitenlohner and D.~Z.~Freedman,
``Stability In Gauged Extended Supergravity,''
Annals Phys.\  {\bf 144}, 249 (1982).}

\nref\BH{ I.~Brunner and K.~Hori, ``Orientifolds and Mirror
Symmetry,'' [arXiv:hep-th/0303135].}

\nref\horietal{ I.~Brunner, K.~Hori, K.~Hosomichi and J.~Walcher, ``Orientifolds
of Gepner Models,''
[arXiv:hep-th/0401137].}

\nref\BrusteinXN{
R.~Brustein and S.~P.~de Alwis,
``Moduli potentials in string compactifications with fluxes: Mapping the
discretuum,''
[arXiv:hep-th/0402088].}

\nref\BurgessIC{
C.~P.~Burgess, R.~Kallosh and F.~Quevedo,
``de Sitter String Vacua from Supersymmetric D-Terms,''
JHEP {\bf 0310}, 056 (2003)
[arXiv:hep-th/0309187].
}

\nref\chsw{P.~Candelas, G.~Horowitz, A.~Strominger, and E.~Witten,
`` Vacuum Configurations for Superstrings,'' \np\ B {\bf 258}, 46 (1985).}

\nref\candelas{P.~Candelas, X.~C.~de la Ossa, P.~S.~Green and
L.~Parkes, ``A Pair of Calabi-Yau Manifolds as an Exactly Soluble
Superconformal Theory,'' \np\ B {\bf 359}, 21 (1991).}

\nref\candtwo{
P.~Candelas, A.~Font, S.~Katz and D.~R.~Morrison,
``Mirror Symmetry for Two Parameter Models. 2,'' \np\ B {\bf 429}, 626 (1994)
[arXiv:hep-th/9403187].}

\nref\cardoso{G.~L.~Cardoso, G.~Curio, G.~Dall'Agata, D.~Lust,
P.~Manousselis and G.~Zoupanos, ``Non-Kaehler String Backgrounds
and their Five Torsion Classes,'' \np\ B {\bf 652}, 5 (2003)
[arXiv:hep-th/0211118].}

\nref\casas{J.A. Casas, Z. Lalak, C. Munoz and G.G. Ross,
Nucl. Phys. {\bf B347} (1990) 243.}

\nref\CKYZ{T.-M. Chiang, A. Klemm, S.-T. Yau and E. Zaslow, ``Local Mirror
Symmetry: Calculations and Interpretations'',
\atmp\ {\bf 3}, 495 (1999) [arXiv:hep-th/9903053].}

\nref\cohen{H.~Cohen,
{\it A Course in Computational Algebraic Number Theory},
Springer-Verlag (1993).}

\nref\DasguptaIJ{
K.~Dasgupta and S.~Mukhi,
``F-Theory at Constant Coupling,'' \pl\ B {\bf 385}, 125 (1996)
[arXiv:hep-th/9606044].}

\nref\dasgupta{K.  Dasgupta, G. Rajesh and S. Sethi, ``M Theory, Orientifolds
and G-Flux,'' JHEP
{\bf 9908}, 023 (1999) [arXiv:hep-th/9908088].}

\nref\DAuriaKX{
R.~D'Auria, S.~Ferrara and M.~Trigiante,
``C-Map,Very Special Quaternionic Geometry and Dual K\"ahler Spaces,'' \pl\ B
{\bf 587}, 138 (2004)
[arXiv:hep-th/0401161].}

%

\nref\dd{
F.~Denef and M.~R.~Douglas,
``Distributions of Flux Vacua,''
[arXiv:hep-th/0404116].
}

\nref\ddtwo{F.~Denef and M.~R.~Douglas, to appear.}

\nref\deren{
J.~P.~Derendinger, L.~E.~Ibanez and H.~P.~Nilles,
``On The Low-Energy Limit Of Superstring Theories,''
Nucl.\ Phys.\ B {\bf 267}, 365 (1986).}

\nref\DiaconescuUA{
D.~E.~Diaconescu and S.~Gukov, ``Three Dimensional ${\cal N} = 2$
Gauge Theories and Degenerations of Calabi-Yau Four-Folds,'' \np\ B
{\bf 535}, 171 (1998) [arXiv:hep-th/9804059].  }

\nref\drom{
D.-E.~Diaconescu and C.~Romelsberger,
``D-Branes and Bundles on Elliptic Fibrations,'' \np\ B {\bf 574}, 245 (2000)
[arXiv:hep-th/9910172].}

\nref\dmw{ D.-E. Diaconescu, G. Moore and E. Witten, ``A
Derivation of K-Theory from M-Theory,'' [arXiv:hep-th/0005091].}

\nref\DineRZ{
M.~Dine, R.~Rohm, N.~Seiberg and E.~Witten,
``Gluino Condensation In Superstring Models,''
Phys.\ Lett.\ B {\bf 156}, 55 (1985).}

\nref\dineseiberg{M.~Dine and N.~Seiberg,
``Is the Superstring Weakly Coupled?'',
\pl\ B {\bf 162}, 299 (1985).}

\nref\racetrack{ L.~Dixon, V.~Kaplunovsky and M. Peskin, unpublished;
L.J. Dixon, SLAC-PUB-5229, 1990.}

\nref\DGW{R. Donagi, A. Grassi and E. Witten, ``A Non-Perturbative
Superpotential with $E_8$ Symmetry'',
{\it Mod. Phys. Lett.} A {\bf 11}, 2199 (1996) [arXiv:hep-th/9607091].}

\nref\DougStr{M. R. Douglas,
``D-Branes and $N=1$ Supersymmetry,''
in the proceedings of Strings 2001, Mumbai, India.
[arXiv:hep-th/0105014].}

\nref\jhstalk{M. R. Douglas, Lecture at JHS60, October 2001,
Caltech. Available on the web at {\tt http://theory.caltech.edu}.}

\nref\dgjt{M. R. Douglas, S. Govindarajan, T. Jayaraman and A. Tomasiello, ``
D-Branes on Calabi-Yau Manifolds and Superpotentials'',
[arXiv:hep-th/0203173].}

\nref\stat{M.~R.~Douglas, ``The Statistics of String / M Theory
Vacua,'' JHEP {\bf 0305}, 046 (2003) [arXiv:hep-th/0303194].}

\nref\mrdtalks{M. R. Douglas, talks at Strings 2003 and at the
2003 Durham workshop on String Phenomenology.}

\nref\dsz{M. R. Douglas, B. Shiffman and S. Zelditch,
``Critical Points and Supersymmetric Vacua,''
[arXiv:math.CV/0402326].}

\nref\fischler{
W.~Fischler, A.~Kashani-Poor, R.~McNees and S.~Paban,
``The Acceleration of the Universe, a Challenge for String Theory,''
JHEP {\bf 0107}, 003 (2001)
[arXiv:hep-th/0104181].}

\nref\FreedVC{
D.~S.~Freed and E.~Witten,
``Anomalies in string theory with D-branes,''
[arXiv:hep-th/9907189].}

\nref\frey{A. R. Frey and J. Polchinski, ``N=3 Warped
Compactifications,'' \pr\ D {\bf 65}, 126009 (2002) [arXiv:hep-th/0201029].}

\nref\fulton{W.~Fulton, ``{\it Introduction to Toric Varieties,}''
Princeton University Press (1993).}

\nref\gkp{S.~B.~Giddings, S.~Kachru and J.~Polchinski,
 ``Hierarchies from Fluxes in String Compactifications,''
\pr\ D {\bf 66}, 106006 (2002) [arXiv:hep-th/0105097] }

\nref\giddings{S.~B.~Giddings, ``The Fate of Four Dimensions'', \pr\ D {\bf 68},
026006 (2003)
[arXiv:hep-th/0303031].}

\nref\GiddingsVR{
S.~B.~Giddings and R.~C.~Myers,
``Spontaneous decompactification,''
[arXiv:hep-th/0404220].}

\nref\gimonpolchinski{E. Gimon and J. Polchinski, ``Consistency Conditions for
Orientifolds and D-Manifolds,''
\pr\ D {\bf 54}, 1667 (1996), [arXiv:hep-th/9601038].}

\nref\GiryavetsVD{ A.~Giryavets, S.~Kachru, P.~K.~Tripathy and
S.~P.~Trivedi, ``Flux Compactifications on Calabi-Yau
Threefolds,'' [arXiv:hep-th/0312104].}

\nref\goheer{N. Goheer, M. Kleban and L. Susskind,
``The Trouble with de Sitter Space,''
JHEP 0307 (2003) 056,
[arXiv:hep-th/0212209].}

\nref\gv{R.~Gopakumar and C.~Vafa, ``On the Gauge Theory/Geometry
Correspondence,'' \atmp\  {\bf 3}, 1415 (1999)
[arXiv:hep-th/9811131].}

\nref\gp{M.~Gra\~na and J.~Polchinski, ``Gauge/Gravity Duals with
Holomorphic Dilaton,'' \pr\ D {\bf 65}, 126005 (2002)
[arXiv:hep-th/0106014].}

\nref\G{A. Grassi,
``Divisors on Elliptic Calabi-Yau 4-Folds and the Superpotential in F-Theory --
I'',
[arXiv:math.AG/9704008].}

\nref\GreeneUD{
B.~R.~Greene and M.~R.~Plesser,
``Duality in Calabi-Yau Moduli Space,'' \np\ B {\bf 338}, 15 (1990).}

\nref\GL{T.W.~Grimm and J.~Louis, ``The Effective Action of ${\cal
N}=1$ Calabi-Yau Orientifolds'', [arXiv:hep-th/0403067].}

\nref\gvw{S.~Gukov, C.~Vafa and E.~Witten, ``CFT's from Calabi-Yau
Four-folds,'' \np\ B {\bf 584}, 69 (2000) [Erratum-ibid.\
B {\bf 608}, 477 (2001)] [arXiv:hep-th/9906070].}

\nref\gukov{S.~Gukov, ``Solitons, Superpotentials and
Calibrations,'' \np\ B {\bf 574}, 169 (2000)
[arXiv:hep-th/9911011].}

\nref\gurrieri{S.~Gurrieri, J.~Louis, A.~Micu and D.~Waldram,
``Mirror Symmetry in Generalized Calabi-Yau Compactifications,'' \np\
B {\bf 654}, 61 (2003) [arXiv:hep-th/0211102].}

\nref\horne{J.~H.~Horne and G.~W.~Moore, ``Chaotic Coupling
Constants,'' \np\ B {\bf 432}, 109 (1994) [arXiv:hep-th/9403058].}

\nref\HullTown{C.~M. Hull and P.~K. Townsend, "Unity of Superstring Dualities,"
\np\ B {\bf 438}, 109 (1995) [arXiv:hep-th/9410167].}

\nref\Ii{V.A. Itskovskih, ``Fano 3-folds I'', {\it Izv. Akad. Nauk.} (Engl.
trans. {\it Math USSR}, Izv {\bf 11}) {\bf 41} 1977.}

\nref\Iii{V.A. Itskovskih, ``Fano 3-folds II'', {\it Izv. Akad. Nauk.} (Engl.
trans. {\it Math USSR}, Izv {\bf 12}) {\bf 42} 1977.}

\nref\kklt{S.~Kachru, R.~Kallosh, A.~Linde and S.~P.~Trivedi, ``de
Sitter Vacua in String Theory,'' [arXiv:hep-th/0301240].}

\nref\kst{S.~Kachru, M.~B.~Schulz and S.~Trivedi, ``Moduli
Stabilization from Fluxes in a Simple IIB Orientifold,''
[arXiv:hep-th/0201028].}

\nref\kachru{S.~Kachru, M.~B.~Schulz, P.~K.~Tripathy and
S.~P.~Trivedi, ``New Supersymmetric String Compactifications,''
JHEP {\bf 0303}, 061 (2003) [arXiv:hep-th/0211182].}

\nref\kknew{S. Kachru and S. Trivedi, private communication and
work in progress.}

\nref\katzvafa{S.~Katz and C.~Vafa,
``Geometric Engineering of ${\cal N} = 1$ Quantum Field Theories,'' \np\ B
{\bf 497}, 196 (1997) [arXiv:hep-th/9611090].}

\nref\KlebanovHB{
I.~R.~Klebanov and M.~J.~Strassler,
``Supergravity and a confining gauge theory: Duality cascades and
chiSB-resolution of naked singularities,''
JHEP {\bf 0008}, 052 (2000)
[arXiv:hep-th/0007191].}

\nref\KLRY{A. Klemm, B. Lian, S.-S. Roan and S.-T. Yau,
``Calabi-Yau Fourfolds for M- and F-Theory Compactifications'',
\np\ B {\bf 518}, 515 (1998) [arXiv:hep-th/9701023].}

\nref\krasnikov{N.~V.~Krasnikov, \pl\ B {\bf 193}, 37 (1987).}

\nref\kreuzer{M. Kreuzer and H. Skarke, \atmp\ {\bf 4}, 1209 (2002)
[arXiv:hep-th/0002240].}

\nref\KSi{M. Kreuzer and H. Skarke, {\tt
http://hep.itp.tuwien.ac.at/kreuzer/CYhome.html}.}

\nref\lu{Z. Lu, private communication;
Z.~Lu and M.~R. Douglas, work in progress.}

\nref\Mayr{P. Mayr,
``Mirror Symmetry, ${\cal N}=1$ Superpotentials and Tensionless Strings on
Calabi-Yau Forfolds,'' [arXiv:hep-th/9610162].}

\nref\M{K. Mohri, ``F-Theory Vacua in Four Dimensions and Toric Threefolds'',
{\it Int. J. Mod. Phys.} A {\bf 14}, 845 (1999) [arXiv:hep-th/9701147].}


\nref\mooreK{
G.~Moore,
``K-theory from a Physical Perspective,'' [arXiv:hep-th/0304018].}

\nref\MooreFG{ G.~W.~Moore, ``Les Houches Lectures on Strings and
Arithmetic,'' [arXiv:hep-th/0401049].}

\nref\greg{ G.~W.~Moore, private communication and work in progress.}

\nref\MMi{S. Mori and S. Mukai, ``Classification of Fano 3-Folds with $B_2\geq
2$'',
{\it Man. Math.} {\bf 36}, 147 (1981).}

\nref\MMii{S. Mori and S. Mukai, ``On Fano 3-Folds with $B_2\geq 2$'',
{\it Adv. Stud. Pure Math.} {\bf 1}, 101 (1983).}

\nref\psflux{J. Polchinski and A. Strominger, \pl\ B {\bf 388}, 736
[arXiv:hep-th/9510227].}

\nref\SenYI{
A.~Sen,
``Dyon - monopole bound states, selfdual harmonic forms on the
multi - monopole moduli space, and SL(2,Z) invariance in string theory,''
Phys.\ Lett.\ B {\bf 329}, 217 (1994) [arXiv:hep-th/9402032].}

\nref\Si{A. Sen, ``Orientifold Limit of F-Theory Vacua'',
\pr\ D {\bf 55}, 7345 (1997) [arXiv:hep-th/9702165].}

\nref\RobbinsHX{
D.~Robbins and S.~Sethi,
``A barren landscape,''
[arXiv:hep-th/0405011].}

\nref\strom{A.~Strominger, ``Superstrings with Torsion'',
\np\ B {\bf 274}, 253 (1986).}

\nref\stromspec{A.~Strominger, ``Special Geometry,'' \cmp\ {\bf
133}, 163 (1990).}

\nref\susskind{L.~Susskind, ``The Anthropic Landscape of String Theory,''
[arXiv:hep-th/0302219].}

\nref\vafaF{ C.~Vafa,
``Evidence for F-Theory,'' \np\ B {\bf 469}, 403 (1996)
[arXiv:hep-th/9602022].}

\nref\WitStr{E.~Witten, "String Theory Dynamics in Various Dimensions,"
\np\ B {\bf 443}, 85 (1995) [arXiv:hep-th/9503124].}

\nref\W{E.~Witten, ``Non-Perturbative Superpotentials in String
Theory'', \np\ {\bf B474}, 343 (1996) [arXiv:hep-th/9604030].}

%
%

\newsec{Introduction}

The problem of stabilizing moduli in superstring compactification has
seen much recent progress.  A benchmark in this progress is the work
of Kachru, Kallosh, Linde and Trivedi \kklt, which builds on many
works on Calabi-Yau compactification, flux compactification and other
aspects of string compactification to propose a way to construct
$\CN=1$ and non-supersymmetric vacua with all moduli stabilized in a
controlled regime.  In this paper, we look for specific models of the
type they suggest, announce examples which we expect will work in
all detail, and explain various senses in which these examples are
less common than one might have supposed.

We begin with a short review of the problem.  Soon after the
pioneering works on Calabi-Yau compactification of the heterotic
string \chsw, it was realized that a disadvantage of this construction
was the presence of moduli of the Ricci-flat metrics and vector
bundles which it uses.  At least in perturbation theory, these moduli
and the dilaton become massless scalar fields with at least
gravitational strength couplings, whose presence would be in
contradiction with experiment.

One might think that this consideration prefers other
compactifications with no moduli.  However, another possibility is
that the problem is just an artifact of perturbation theory.  One can
see this by postulating simple effective potentials which could
plausibly emerge from non-perturbative effects in string theory, which
have isolated minima.  Any of these minima would be a vacuum with
stabilized moduli.

Some early examples were the models of \refs{\DineRZ,\deren},
and the ``racetrack'' models
\refs{\racetrack,\krasnikov,\casas}\ which use non-perturbative contributions
from two gauge theory sectors, with different dependences on the gauge
coupling.  Many similar constructions have been proposed over the
years.  Their essential ingredient, more than any particular feature
of string theory, is the concept of effective potential, in which
effects from various different sources are added into a single
function which controls the vacuum structure.  Once one believes that
this is an accurate picture of the situation, then given a
sufficiently rich supply of contributions to the potential, it becomes
very plausible that the typical potential will have many isolated
minima, so that moduli stabilization is generic.  Furthermore, the
value of the effective potential at the various minima (the
``effective cosmological constant'') will be widely distributed among
both positive and negative values, possibly including the observed
value.

One might ask whether other general features of the problem, such as
low energy supersymmetry, change this general expectation.  As we have
discussed at length elsewhere \dd, the standard expression for
the potential from $\CN=1$ supergravity, supports this point of view.
By postulating a superpotential and K\"ahler potential, one can get a
wide range of potentials, which do not obviously favor positive or
negative effective cosmological constant.

One important general expectation from string theory is the following.
Calabi-Yau compactification, and essentially any ``geometric''
compactification of string and M theory, has a ``large volume limit''
which approaches ten or eleven dimensional Minkowski space.  In this
limit, the four dimensional effective potential will vanish
\dineseiberg.  Recently more general arguments for this claim were made
by Giddings \giddings.  This makes it somewhat harder to
stabilize vacua with zero or positive effective cosmological constant,
as one needs the potential to have at least two points of inflection
(the desired minimum, and a maximum or more generally a barrier at
larger compactification radius).  Still, this leads to no obvious
problem of principle.

To go beyond these rather general claims, one must get sufficient
control over nonperturbative effects to show that such potentials
could arise from string/M theory, and be able to compute them in a
wide enough range of examples to make serious predictions.
This line of thought developed in the late 1980's, and led to a focus
on nonperturbative methods in string theory, a field which took off
with the 1989-90 work on matrix models.  An equally important
early nonperturbative result was the exact solution, including all
world-sheet instanton corrections, of $(2,2)$ Calabi-Yau sigma models
using mirror symmetry \candelas.

While when first formulated these techniques appeared to address only
low dimensional models, the 1993-94 work of Seiberg and collaborators
on nonperturbative supersymmetric gauge theory showed that similar
results could be obtained for the effective superpotential in simple
four dimensional models.  With the 1994-95 realization
\refs{\SenYI,\HullTown,\WitStr} that duality and branes were key
nonperturbative concepts, tremendous advances were made, leading to a
fairly good picture of compactifications with at least $8$
supercharges, and some success in describing compactifications with
less or no supersymmetry.

Studying such compactifications and mapping out the possibilities
requires both the techniques to do Calabi-Yau, F theory, $G_2$
compactification etc. and study the physics of each, and
``classification'' and ``mapping'' results, such as the classification
of of Calabi-Yau's which are toric hypersurfaces, and the detailed
study of their moduli spaces, for example \kreuzer.  These should both
provide specific examples of these general recipes and some idea of
which ones might be relevant for describing the real world.  We will
call upon a number of these results below
\refs{\G,\Ii,\Iii,\KLRY,\MMi,\MMii}.

The subject of compactification with four or fewer supercharges
remains somewhat controversial.  Most works on this subject freely use
field theoretic concepts such as Kaluza-Klein reduction and the
effective potential, with the main string/M theory inputs being new
light states and new corrections to the effective Lagrangian.  At
present, there is no real evidence that this is wrong.

On the other hand, it is true (by definition) that a vacuum with
stabilized moduli is not connected to the limit of arbitrarily weak
coupling and arbitrarily large compactification volume through
time-independent solutions.  One must pass either through the larger
configuration space (``go off-shell'' in an old-fashioned language) or
consider time-dependent solutions to do this, and our lack of
understanding of these subjects in string/M theory makes it conceivable
that what look like valid solutions from an effective field theory
point of view in fact are not valid solutions of string/M theory.
This possibility has been put forward most forcefully by Banks and
collaborators \refs{\banks,\BanksES}.
While we do not find their present arguments
convincing, the point is definitely not settled and deserves attention.

Another issue along these lines is the question of whether there
is some obstacle within string/M theory to realizing compactifications
with positive vacuum energy.  Over the mid-90's, convincing observational
evidence was found for an accelerated expansion of the universe,
which is most simply explained by the hypothesis that there
is a non-zero positive ``dark energy'' of energy density about $0.7$
times the critical value.  While there are many possibilities for what
this is, the simplest and probably easiest to reproduce from string
theory is a cosmological constant, which would appear to cause the
long term evolution of the universe to asymptote to a de Sitter
geometry.  From the point of view of field theory and the effective potential,
there is no problem with getting positive vacuum energy.
On the other hand, no such string vacuum is known, and
theoretical arguments were even advanced that these should not exist
\refs{\banks,\fischler,\goheer}.

Meanwhile, progress in our understanding of string/M theory has led to
steady progress in the study of moduli stabilization.  An important
recent work in this direction is that of Kachru, Kallosh, Linde and
Trivedi \kklt, who argue that metastable de Sitter vacua with all
moduli stabilized can be constructed in type
\IIb\ string theory, at least in an effective potential framework.

The KKLT work has two aspects.  In the first, they
propose a recipe for such a construction, which we will discuss in
detail shortly.  In the second, they sidestep the claim that
de Sitter space could not be realized in string theory, by
suggesting that our universe need not asymptote to de Sitter
space.  Rather, our vacuum might be metastable, in a positive local
minimum of the effective potential.  As we dicussed, since the
potential goes to zero at large compactification volume, any such
vacuum is potentially metastable to decay to this decompactification
limit.  Using results of Coleman and de Luccia, Hawking and Moss, and
others on tunneling in semiclassical quantum gravity, KKLT argued that
for almost any reasonable potential, one would find an extremely small
but non-zero decay rate, slow enough to pose no cosmological
constraint, but fast enough to evade the paradoxes which had been
suggested might forbid eternal de Sitter space as a vacuum.
These ideas are reviewed in \bala\ (see also the recent \GiddingsVR).

This leaves the more technical problem of showing that such metastable
de Sitter vacua actually do exist in string theory compactification.
KKLT suggested a recipe by which this could be done,
following previous work of many authors
on moduli stabilization by fluxes (a very incomplete list includes
\refs{\achflux,\beckerG,\boupol,\cardoso,\dasgupta,\frey,%
\gp,\gurrieri,%
\kst,\kachru,\psflux,\strom}) and especially Giddings,
Kachru and Polchinski \gkp\ who developed a class of \IIb\ flux
compactifications which stabilize the dilaton and complex structure
moduli.  Some advantages of this construction are that it allows
fixing the dilaton to weak coupling by making an appropriate choice of
flux, and that cancellations can bring the flux
contribution to the potential below the string scale.  Most
importantly, it is based on the relatively well understood theory of
Calabi-Yau moduli spaces, so explicit computations can be done.

Having this construction in hand, the remaining problem is to
stabilize the K\"ahler moduli and obtain a small positive cosmological
constant.  Granting the effective potential framework, one proceeds as
in the early works on moduli stabilization, to look for
nonperturbative contributions to the superpotential which could
stabilize the remaining moduli.  As it happens, in \IIb\
compactification nonperturbative effects from gauge theories on
branes, and from D-instantons, all depend on K\"ahler moduli, and thus
one would expect that generically this stabilization would not be a
problem -- once one had computed these constributions to the effective
potential sufficiently well, they would stabilize all moduli.

Finally, to obtain a positive cosmological constant, one has several
choices.  Conceptually the most straightforward is to look for
supersymmetry breaking vacua of the previously computed effective
potential, which again on grounds of genericity should exist.
Alternatively, one can look for supersymmetric AdS vacua which
stabilize all moduli, and then add other supersymmetry breaking
effects which lift the vacuum energy.  This idea has the advantage
that the resulting potential can naturally have the barrier required
for metastability of a de Sitter vacuum.  KKLT suggested to add anti
D$3$-branes, and one can imagine many other possibilities such as D
terms coming from brane sectors \refs{\DougStr,\BurgessIC,\BinetruyHH}.

The upshot is that such effects, if they exist, could well combine to
stabilize all moduli at a metastable de Sitter minimum.  There are two
obvious gaps in this argument.  The deeper one is the question of
whether the effective potential analysis is valid.  Perhaps the best
way to address this is to consider a specific example, and try to
justify all of its ingredients in ten dimensional string theory.

To do this, one must address the first gap, which is that no
concrete model of this type has yet been put forward.
In this work, we will fill this gap, by examining models of the
general class discussed by KKLT, to see if this idea can
actually be implemented in string theory.  We will need to call upon
many of the ideas discussed above, especially the classification
of Calabi-Yau three-folds and four-folds, and the analysis of
nonperturbative effects in F theory by Grassi \G.

While we will exhibit models in which nonperturbative effects lift all
K\"ahler moduli, we will also argue that, at least if we rely on
instanton effects, such models are {\it not} generic.  There is a
fairly simple reason to expect this.  In the language of orientifold
compactification with branes, it is that these world-volume gauge
theories usually have too much matter to generate superpotentials,
as seen in works such as \BH, and as we will see in a large class of models
below.
This is because the cycles on which the branes wrap can be deformed,
or because the branes carry bundles with moduli.  In either case, the
gauge theory has massless adjoint matter, which eliminates the
superpotential.  The simplest case is the $0$-cycle or point, which
clearly always has moduli.  Of the two-cycles, only $S^2$'s can be
rigid; higher genus Riemann surfaces in Calabi-Yau almost always come
with moduli.  Bundles on surfaces (and the entire CY) normally have
moduli as well.

On the other hand, in a sizable minority of models, the instanton
generated superpotentials are sufficiently generic to stabilize all
K\"ahler moduli.  The examples that we can check systematically are
toric fourfolds $X$ elliptically fibered over a three dimensional base
$B$ which is Fano, and $\IP^1$-bundles over a toric surface.
Much of what we say should hold more generally.

There are no very simple models in this class.  In particular, one can
argue that no model with one K\"ahler modulus can work.\foot{
This conclusion was reached independently by D. Robbins
and S. Sethi \RobbinsHX.
Also, relevant work of V. Balasubramanian and P. Berglund is mentioned
in \bala.}

Of the models which work, we have considered three of the simpler ones
in some detail. The fourfolds are elliptic fibrations over the Fano
threefolds ${\cal F}_{11}$ and ${\cal F}_{18}$
\refs{\MMi,\MMii,\M}. Their Euler characteristics are $\chi=16848$ and
$\chi=13248$ respectively. Some of our considerations will require
that we work with their orientifold limits.  These are Calabi-Yau
threefolds that can be realized as hypersurfaces in toric varieties
and their Hodge numbers are $h^{1,1}=3,h^{2,1}=111$ and
$h^{1,1}=5,h^{2,1}=89$ respectively.  Finally, we consider an
elliptic fibered CY over $\IP^2$, a much studied CY \candtwo\ with
$h^{1,1}=2$ and $h^{2,1}=272$.

\newsec{Review of KKLT construction}

We start with \IIb\ superstring theory compactified on a Calabi-Yau
orientifold, specified by a choice of CY threefold $Z$ and a holomorphic
involution $\hat\Omega$ on $Z$ whose fixed locus consists of points and
surfaces (O$3$ and O$7$ planes).
For example, we might take $Z$ to be the hypersurface in $\BC\BP^4$
defined by a quintic polynomial $f(z)=0$, and $\hat\Omega$ to be
the map $z_1\rightarrow -z_1$, $z_i\rightarrow z_i$ for $i>1$.

We then introduce D-branes in such a way as to cancel the RR tadpoles
produced by the orientifold planes, as in \gimonpolchinski.
For a recent discussion of this problem, see \horietal.

The number of complex structure moduli of $Z$ is $h^{2,1}(Z)$, and
the number of complexified K\"ahler moduli is $h^{1,1}(Z)$.
We are going to refer to these as ``shape'' and ``size'' moduli
respectively, mostly because the words ``complex'' and ``K\"ahler''
have so many different roles in the discussion that the usual
terminology is cumbersome (one gets tired of saying ``K\"ahler metric
on the K\"ahler moduli space,'' etc.)

A more general starting point would be F theory \vafaF.  We will use this
language to construct models below, but rapidly move to the \IIb\
orientifold limit in our examples, since there are many unanswered
physical questions in either picture, which we will need to appeal
to the underlying definitions in string theory to resolve.

The basic relation is as follows.  An F theory compactification is
defined by a choice of Calabi-Yau fourfold $X$ with an elliptic fibration
structure over a threefold $B$,
\eqn\ellfiber{\xymatrix{
  T^2     \ar[r] &  X \ar[d]^{\pi}\cr
  & B.\cr
}}
Physically, we compactify \IIb\ theory on $B$, introducing $7$-branes
at the singularities of the fibration $\pi$, so that the resulting
dilaton-axion field at a point $p\in B$ corresponds to the complex
structure modulus $\tau$ of the fiber $\pi^{-1}(p)$.
The orientifold limit \Si\ is then the special case in which
all of these singularities are $D_4$ singularities (an O$7$-plane and
four coincident D$7$'s).  In this case, $Z$ is a double cover of $B$
branched at the singularities, and $\hat\Omega$ exchanges the two sheets.
Let us denote the part of the cohomology of $Z$ even or odd under $\hat\Omega$
with subscripts $\pm$.

Classically and in the absence of fluxes, this compactification
has a moduli space of $\CN=1$ supersymmetric vacua, parametrized
by $h_-^{2,1}$ shape moduli, the dilaton-axion,
$h_+^{1,1}$ size moduli complexified by the RR 4-form
potential $C_4$, and $h_-^{1,1}$ moduli from the 2-form
potentials $B_2$ and $C_2$ \refs{\BH,\GL}.

We will avoid having to discuss the 2-form moduli by restricting
attention to models with $h_-^{1,1}=0$.  More generally, while we do
not know a mechanism which would stabilize them at large volume, they
might be stabilized by world-sheet and D$1$-instantons, or by
couplings to brane world-volumes.

In general, there are also open string moduli corresponding to
positions of D7-branes and D3-branes, and moduli of bundles on
D7-branes.  We will ignore these through most of the discussion and
return to them in the conclusions.  One excuse for this is that these
moduli spaces are all expected to be compact.  This is intuitively
clear for positions of D3 branes, since their moduli space is $Z$
itself, which is compact.  This should also be true for moduli spaces
of bundles except for small instanton limits, but these just
correspond to other brane configurations.  Thus, there is no analog of
the decompactification or runaway problems for these moduli, and
generic corrections to the potential should fix them.

\subsec{Fixing moduli using fluxes}

We begin by trying to fix the shape moduli and the dilaton-axion.  In
\IIb\ language, this is done by turning on the NS and RR three-form
field strengths, $H^{(3)}$ and $F^{(3)}$ respectively.  As discussed
in many references (e.g. \refs{\gkp,\ad}) the equations of motion will
force these to be harmonic forms, so they are determined by their
cohomology classes in $H^3(Z,\BR)$.  There is a Dirac-type
quantization condition which normally forces these classes be integer
quantized in units of the string scale.  Setting this unit to one,
they live in $H^3(Z,\BZ)$.

The choice of flux is constrained by the tadpole condition on the
RR four-form potential,
 \eqn\tadpole{
 L \equiv {1 \over 2} \int G_4 \wedge G_4 = {1 \over 2} N_R^\alpha
 \, \eta_{\alpha\beta} \, N^\beta_{NS} = {\chi(X) \over 24} - N_{D3},
 }
where $N_{D3}$ is the number of D3-branes minus the number of
anti-D3 branes.

The ten-dimensional
\IIb\ supergravity analysis of unbroken $\CN=1$ supersymmetry
in this context is done in \gp; the most important condition is that
$$
G^{(3)} = F^{(3)} - \tau H^{(3)}
$$
must be imaginary self-dual, $G=i*G$.  One can restate this as the
condition that
$$
G^{(3)} \in H^{(0,3)}(Z,\BC) \oplus H^{(2,1)}(Z,\BC) .
$$
Setting the remaining parts of the cohomology to zero
amounts to $h^{2,1}+1$ conditions, which is precisely enough to
fix all shape moduli and the dilaton.  Thus, we might
expect generic choices of flux to do this.

There is an additional condition for this to be true with four noncompact
Minkowski dimensions: the $H^{(0,3)}$ component of $G^{(3)}$ must be zero.
This is one more condition than the number of moduli, so this is non-generic.
Otherwise, we get supersymmetric AdS vacua.

A nice physical summary of these results, which will be essential later
on, is that these conditions on $G^{(3)}$ follow from solving the
supersymmetry conditions in an
$\CN=1$ effective supergravity theory.  Its configuration space is
the combined Calabi-Yau and dilaton-axion moduli space; in other words
the product of the shape moduli space $\CM_c(Z)$, the
size moduli space $\CM_K(Z)$, and the upper half
plane $\CH$, with the K\"ahler potential
\eqn\lvkahler{
{\cal K} = -\log\Im\tau - \log\int_Z \Omega\wedge\bar\Omega - 2 \log V
}
where $\tau$ is the dilaton-axion,
$V$ is the volume of the Calabi-Yau as a function of the
size moduli (more on this below), and $\Omega$ is the
holomorphic three-form on $Z$.  This is just the K\"ahler potential
of the compactification with zero flux and $\CN=2$ supersymmetry,
computed in the $\alpha',g_s\rightarrow 0$ limit.

As superpotential, we take the Gukov-Vafa-Witten superpotential
\eqn\fluxW{
W_{GVW} = \int \Omega\wedge G^{(3)} .
}
One can see \refs{\gkp,\ad} that the conditions
\eqn\kvev{
0 = D_i W = \p_i W + (\p_i {\cal K}) W
}
for the shape moduli set the $H^{(1,2)}$ part of $G^{(3)}$
to zero, and for the dilaton-axion sets the $H^{(3,0)}$ part to zero.

For typical values of the flux, this will fix $\tau$ and the shape
moduli $z$ at a mass scale $m_0 \sim \alpha'/\sqrt{V}$, where $V$ is
the volume of $Z$.  The typical value of $e^{\cal K}|W|^2$ at the minimum is
$1$ in string units, but this can be made smaller by tuning the
fluxes.  One expects its smallest value to be $\sim 1/{\cal N}_{vac}$
where ${\cal N}_{vac}$ is the number of flux vacua on the complex
structure moduli space, as we discuss below and in \dd.

We will discuss the technicalities of finding choices of flux which
stabilize these moduli in a desired region of moduli space in section 4.
It will turn out that this is often, but not always possible.

\subsec{Fixing size moduli}

Since \fluxW\ is independent of the size moduli, the conditions $D_i
W_{GVW}=0$ for the size moduli amount to either $\p_i {\cal K}=0$ for all
size moduli, or else $W=0$.  It is easy to see (as we discuss below)
that in the large volume limit, one cannot solve $\p_i {\cal K}=0$ for all
moduli.  One can find Minkowski supersymmetric solutions with $W_{GVW}=0$.
One can also find Minkowski but non-supersymmetric vacua with $W_{GVW}\ne 0$,
because of the no-scale structure of \lvkahler.

In either case, while we might find other effects which lead to a
positive vacuum energy, this energy will typically decrease with
increasing compactification volume, and the resulting solutions will
be unstable.  We need a more complicated effective potential to fix
this problem, which might be obtained by incorporating stringy and
quantum corrections to $W$ and ${\cal K}$.

Unfortunately doing this remains a hard problem, both to get results
in examples, and even to define what one means by the nonperturbative
effective potential in general.  In particular, there are essentially
no results on nonperturbative corrections to the K\"ahler potential.

To address this, one can seek examples where the instanton expansion
appears to be valid, so that the leading corrections will dominate.
This was the primary goal in KKLT's discussion and we will address it
below.  However we believe that there is no question of principle
which requires restricting attention to these examples; rather we do
this because of our limited technical control over the theory at
present.

KKLT suggested to stabilize the size moduli using quantum corrections
to the superpotential, as they are known to be present and depend on
the size moduli in many examples, and are controlled at large volume
and weak coupling.

For example, one can consider a case in which D$7$-branes wrapped on a
cycle $\Sigma$ have pure $SU(N)$ Yang-Mills theory as their world-volume
theory.  The gauge coupling in this theory is
${1\over g^2} = V(\Sigma)$
where $V(\Sigma)$ is the volume of the cycle $\Sigma$ in string units,
and thus
nonperturbative effects in this theory lead to a superpotential
 \eqn\Weff{
 W_{eff} = B \, e^{-V(\Sigma)/N}
 }
where $B$ is a (presumably order $1$) quantity determined by threshold
effects, etc. and which may also depend on size and shape moduli.

If we consider a vacuum in which the flux contribution to the
superpotential \fluxW\ takes a value $W_0$ independent of size moduli,
and grant that this is the leading correction, the supersymmetry condition
for the size modulus becomes (let $\rho=i V$),
\eqn\fixrho{
0 = D_\rho W = -{B\over N} e^{i\rho/N}
 - {3\over\Im\rho} \left(W_0 + B e^{i\rho/N}\right) .
}
For small $W_0$, this will normally have a unique solution at large
imaginary $\rho$, roughly given by
$$
\rho \sim -i \log {3 N W_0\over B} .
$$
To see this without detailed calculation, note that the problem of
solving $D_\rho W=0$ with $W\ne 0$ is the same as the problem of
finding critical points of the function $f=e^{\cal K}|W|^2$.  This function
is positive and goes to zero at very large $\Im\rho$.  If we contrive
$W$ to have a zero at large $\Im\rho$, then it is easy to see that as
we increase $\Im\rho$ from this value at fixed $\Re\rho$, we are
always at a critical point in $\Re\rho$.  Furthermore, as we increase
$\Im\rho$, $f$ starts out increasing, so for it to later decrease it
must pass through a critical point in $\Im\rho$.

This solution will have $W\sim W_0$
and thus we will end up with a supersymmetric AdS solution with small
negative cosmological constant, controlled by the previous step where
we chose fluxes to get small $W_0$.

Another source of similar non-perturbative contributions to the
superpotential is D$3$-brane instantons wrapped on surfaces in $Z$ \W.
Their dependence on the size moduli is the same and thus this level of
the discussion works the same way.  As we discuss below, in the case
that D$7$-branes wrap a cycle $\Sigma$, these are equivalent to gauge
theory instantons, but are present more generally.  Thus in the
detailed discussion below, we consider the non-perturbative
corrections as generated by D$3$-brane instantons.

Some further general observations on this problem appear in \BrusteinXN.

\subsec{Breaking supersymmetry}

We will have enough trouble trying to realize the first two steps, but
given this, to go on to the third step we would need some approximate
expression for the warp factor on $Z$, so that we could argue that the
increase in vacuum energy produced by adding an anti D$3$-brane could
also be made small.  The simplest way to imagine this working is to
fix the shape moduli near a conifold singularity, and appeal to the
Klebanov-Strassler solution \KlebanovHB\ as an approximate description of the
metric in this region.

While it seems to us that this should generally work, of course there
is room for more subtleties, as we will briefly discuss below.

\newsec{Details of size moduli stabilization}

We seek the conditions on $Z$ which will lead to a superpotential
which stabilize all size moduli significantly above the string scale,
so that $\alpha'$ corrections can be ignored.
It will emerge from considerations below that there are no models with
$h^{1,1}=1$, and thus we consider the multi-moduli case from the
start.  We will usually set $\ell_s = 2 \pi \sqrt{\alpha'}$ to $1$ in
the following.

Let $D_i$ be a basis of divisors on $Z$ (essentially, these are
classes in $H_4(Z,\BZ)$).  The main data we will need about $Z$ are
the triple intersection numbers of the divisor basis, $D_{ijk} = D_i
\cdot D_j \cdot D_k$.  Using Poincar\'e duality, we also let $D_i$
denote the corresponding class in $H^2(Z,\BZ)$.
For CY's which are hypersurfaces in toric varieties,
there are efficient methods for computing these intersection numbers
and the other data we are about to discuss.  We discuss some relevant examples
in section 5 and appendix B.

We define the size moduli by writing the K\"ahler class $J$ of $Z$
as
 \eqn\Jfx{
 J = \sum_i t^i D_i ,
 }
defining a set of real coordinates $t^i$ on the space of K\"ahler
classes.

The classes which actually correspond to K\"ahler metrics are those lying
in the K\"ahler cone, defined by
\eqn\kahlercone{
0 < \int_C J
}
for all holomorphic curves $C$.  The classes of such curves are
called ``effective classes'' and form a cone, the Mori cone.

In fact, to be able to ignore $\alpha'$ corrections, we want all such
areas of curves to be at least $\CO(1)$ in string units.

The natural holomorphic coordinates which appear in \IIb\
orientifold compactification are not the $t_i$, but
are instead the complexified volumes of
divisors, which we denote $\tau_i$,
 \eqn\taui{
 \tau_i = \int_{D_i}
 {1 \over 2} J \wedge J - i \, C_4 .
 }
We denote their real parts as
$$
 V_i = {\rm Re~} \tau_i .
$$

The K\"ahler potential
on this configuration space, neglecting $\alpha'$ and $g_s$ corrections, is
\eqn\kahlersize{
{\cal K}_K = -2 \ln V(\tau,\bar\tau)
}
where $V$ is the volume of $Z$, defined by
\eqn\volexp{
 V = {1 \over 6} J^3 = {1 \over 6} D_{ijk} t^i t^j t^k .
}
We can write this (implicitly) as a function of $\tau_i$ using
\eqn\voli{
\tau_i = {\p V\over\p t_i} = {1 \over 2} D_i J^2 = {1 \over 2}
 D_{ijk} t^j t^k .
}
This change of coodinates
is essentially a Legendre transform on $V$, as discussed
in detail in \DAuriaKX.

\subsec{Instanton superpotentials}

The best studied corrections to the superpotential in F theory are
produced by D3-brane instantons wrapped around divisors.
These take the form
\eqn\npW{ W_{np} = \sum_{\vec{n}} b_{\vec{n}} \, e^{-2\pi \vec{n}
 \cdot \vec{\tau}}
 }
where the $b_{\vec{n}}$ are one-loop determinants on the divisors
$D_{\vec{n}} = n^i D_i$, depending on the complex structure
moduli. At large radius, $2 \pi \tau_i \gg 1$ and the mass scale
of this contribution to $W$ is exponentially suppressed compared
to $m_0$, so to a good approximation we can consider the complex
structure moduli to be fixed by the tree level flux superpotential
$W_0$, and use $W = W_0 + W_{np}$ with constant $W_0$ and
$b_{\vec{n}}$ as effective superpotential for the Kahler moduli.

The most important thing to know about the determinants $b_{\vec{n}}$
is whether they are non-vanishing.  This was studied in \W\ by using
the relation between F theory and M theory compactification on the
four-fold.  In M theory, such a correction comes from a fivebrane
instanton wrapped on the lift ${\cV}_{\vec n}$ of the divisor to
the fourfold.  The F theory limit is the limit in which the area of
the fiber goes to zero, and a divisor ${\cV}$ will contribute in
this limit if it is vertical, meaning that $\pi({\cV})$ is a proper
subset of $B$.

In this formulation, $b_{\vec n}$ includes the determinant of a Dirac
operator on $\cV$, which can have zero modes.  In fact the zero
modes turn out to be equivalent to holomorphic $p$-forms on
${\cal V}_{\vec n}$ tensored with a three-dimensional spinor.  Let
the number of such $p$-forms be $h^{0,p}$.

An instanton which contributes to $W_{np}$ must have two
fermion zero modes (corresponding to the two supersymmetries broken by
the instanton).  Since $h^{0,0}=1$, this
will be true if $h^{0,j}=0$ for all $j=1,2,3$.  On the other hand,
it is possible for other brane couplings (in particular, couplings to
the NS and RR fluxes we used to stabilize other moduli) to lift zero modes
in pairs.  Thus it is possible that $b_{\vec n}\ne 0$ more generally, but
in any case the necessary criterion for this
is that ${\cal V}_{\vec{n}}$ is a divisor of arithmetic genus
$$
\chi({\cal O}_{\cal V}) \equiv \sum_j (-1)^j h^{0,j} = 1.
$$
In this more general case,
it is also possible for variations in complex structure
to bring down or lift zero modes in pairs, so $b_{\vec{n}}$ could have
zeroes at special points in complex structure moduli space, possibly
depending on the choice of flux.

Unfortunately not much more is known about the determinants
$b_{\vec{n}}$.  Note however that they must be sections of the same
line bundle as the holomorphic 4-form $\Omega_4$ on $X$; otherwise the
combined superpotential wouldn't make sense. Let us fix the overall
scale of $\Omega_4$ (and thus $b_{\vec{n}}$) by putting $\int_X
\Omega_4 \wedge \bar{\Omega}_4 = 1$ at the given point $(\tau,z)$. In
this natural normalization, it is reasonable to assume the
$b_{\vec{n}}$ are generically not much smaller or bigger than 1.

The upshot is that divisors whose lift to $X$ have $h^{0,j}=0$ for
all $j=1,2,3$ will contribute to the superpotential, whether or not
flux is turned on and regardless of the values of the other moduli.
The most general class of divisors which can contribute are the
divisors of arithmetic genus $1$; these contributions might vanish for
special values of flux and moduli.\foot{
One might worry about possible
modifications to this analysis if $\cal V$ is spin$_c$ and not spin,
for reasons discussed in \FreedVC.  These can modify the coefficient
$b$ and might even introduce dilaton dependence \greg, but do not affect the
statements made here or our later conclusions.
We thank G. Moore and S. Sethi for communications on this point.

We also note that the analysis in \W\ used a $U(1)$ symmetry of the
normal bundle to the divisor.  If this were broken by the fluxes,
the arithmetic genus $1$ condition might be relaxed \kknew.}

\subsec{Gauge theory superpotentials}

Superpotentials produced by non-perturbative effects in
gauge theories living on D7-branes can also be understood as coming
from divisors of arithmetic genus 1.

Roughly, if one wraps a D$7$ about a
divisor $\Sigma$, the zero size limit of the Yang-Mills instanton in
the resulting world-volume gauge theory is just a D$3$-instanton
wrapped on $\Sigma$.
A detailed discussion is somewhat more complicated. For example, in
pure $SU(N_c)$ super Yang-Mills theory, the instanton has $2N_c$ fermion
zero modes, which for the corresponding D$3$-instanton arise from
$3-7$ open strings.  Thus it does not contribute directly to the
superpotential.  Rather, its effects force gaugino condensation which
leads to a vacuum energy which can be reproduced by a superpotential.
This accounts for the $1/N$ in the exponent of \Weff.

In F-theory, nonabelian gauge symmetries arise from singularities of
the elliptic fibration. For example, an ${\bf A}_{N-1}$ singularity
fibered over a surface $S$ in the base corresponds to an $SU(N)$ gauge
theory living on $S$. If $S$ has $h^{0,1} = h^{0,2} = 0$ and the
singularity does not change over $S$, there will be no additional
matter and we are in the pure $SU(N)$ case just discussed. In
\katzvafa, it was shown that by compactifying on a circle and
going to the dual M-theory picture, the three dimensional
nonperturbative superpotential can be computed as coming from M5 brane
instantons wrapping the exceptional divisors obtained by resolving the
${\bf A}_{N-1}$ singularity. Again, these divisors are of arithmetic
genus 1. The correct four dimensional result is then obtained by
extremizing the three dimensional superpotential and taking the
decompactification limit.

A similar geometrical analysis can be done when in addition
fundamental matter is present \DiaconescuUA. The situation is more
subtle here, but the upshot is that the gauge theory superpotentials
can again be derived from M5 brane instantons wrapped on divisors of
arithmetic genus 1.

Thus, in the known examples, a necessary condition to generate a
nonperturbative superpotential is the presence of a divisor of
arithmetic genus 1 in the fourfold.\foot{Actually, in \W, a more
general possibility is raised: divisors of arithmetic genus $\chi(D)
>1$ (but presumably never $\chi \le 0$) might contribute to the
superpotential, through strong infrared dynamics or ``fractional
instanton'' effects.  We thank S. Kachru for reminding us of this point.}

A more intuitive but only partial argument for the role of the
arithmetic genus in the brane gauge theory is that the cohomology
groups $H^{0,p}({\cal V})$ which entered our previous discussion also
control the matter content on the D$7$-brane gauge theories.  In
particular, non-zero $h^{0,1}$ and $h^{0,3}$ will lead to massless
adjoint matter, which tends to eliminate the superpotential (as in
$\CN=2$ gauge theory).  This conclusion is not strict as gauge theory
with a non-trivial world-volume superpotential for the adjoint matter
(say $\Tr \phi^n$) can generate a superpotential, but this possibility
comes along with the possibility of deforming the superpotential to
give mass to the adjoint matter and thus is accounted for by the
possibility we discussed earlier of lifting the zero modes in pairs
(this can be made more precise by relating the superpotential to
obstruction theory and $h^{0,2}$ as in the D-brane literature
\dgjt).

\subsec{Complete sets of divisors}

Divisors of arithmetic genus $1$ tend to be rare \refs{\W,\G}.
Furthermore, in order to
stabilize the compactification at strictly positive radii, a
sufficient number of distinct divisors must contribute. This can be seen
as follows.

We are looking for solutions to $D_{\tau_i} W =
\partial_{\tau_i} W + (\partial_{\tau_i} {\cal K}) W = 0$.
Now
\eqn\dK{
 d {\cal K}_K = - {1 \over V} J^2 d J
 = - {1 \over 2 V} J d J^2
 = -{t^i dV_i \over V},
 }
hence $\partial_{\tau_i} {\cal K} = - {t^i \over 2 V}$, and
 \eqn\DW{
 0 = D_{\tau_i} W = \sum_{\vec{n}} b_{\vec{n}} \, e^{-2\pi \vec{n}
 \cdot \vec{\tau}} (-n_i) \, \, - {t^i \over 4\pi V} W.
 }
By multiplying with $D_i$ and summing, it follows that
 \eqn\lincombdiv{
 \sum_{\vec{n}} b_{\vec{n}} \, e^{-2\pi \vec{n}
 \cdot \vec{\tau}} D_{\vec{n}} \, + {W \over 4\pi V} J = 0.
 }
We want vacua with $W\neq 0$.  This is
generic, and if the contributing divisors
are linearly independent, as is often the case in examples, it
follows directly from this equation.

In this case, the
K\"ahler class $J$ is a linear combination of the contributing
divisors. Thus,
a necessary condition for a solution with positive radii to exist
is that a linear combination
$$
{\cal R} = r_{\vec{n}} D_{\vec{n}}
$$
of the contributing divisors exists such
that ${\cal R}$ lies within the K\"ahler cone \kahlercone.
We will call such a set
of divisors ``complete.''

Note that a set of divisors cannot be complete if there exists a curve
$C$ which does not intersect any divisor in that set, because then all
${\cal R}$ will lie outside the K\"ahler cone, or at best on its
boundary.  On the other hand a full basis of divisors is automatically
complete. Since it is easy to verify if a set of divisors is a basis,
we will focus mainly on examples in which a full basis in $H_4(Z)$ of
divisors contributes to the superpotential.

\subsec{Parameters}

At large volume, the exponentials in \lincombdiv\ are very small,
so for a solution to exist, $W$ and therefore $W_0$ has to be very
small, of the order of the dominant exponentials $e^{-2 \pi
\vec{n}\cdot\vec{\tau}}$. Again because $J$ in \lincombdiv\ has to
lie well inside the K\"ahler cone, the set of contributing
divisors $D_{\vec{n}}$ which have $e^{-2\pi\vec{n}\cdot\vec{\tau}}
\sim W_0$ has to be complete in the sense introduced above, since
the terms for which $e^{-2\pi\vec{n}\cdot\vec{\tau}} \ll W_0$ can
effectively be neglected.

For a suitable choice of basis of divisors, the $t^i$ in \Jfx\
correspond to curve areas, so a rough estimate for the divisor
volumes is $V_i \sim k A^2$, where $A$ is the average curve area
and $k$ some constant, which increases with the number of K\"ahler
moduli due to the increase in terms in \voli\ (in examples $k$ is
at least proportional to the number of K\"ahler moduli). So an
estimate for the average curve size at a critical point of the
superpotential is
 \eqn\Aest{
 A \sim \biggl({\ln |W_0|^{-1} \over 2 \pi k}\biggr)^{1/2}.
 }

We see from this estimate that for many K\"ahler moduli (so $k$ is
large), $W_0$ has to be exceedingly small to have $A>1$.


Moreover for many moduli the probability that some curves are
significantly smaller than the average increases, making it even
more difficult to fix the moduli in a controlled regime. This
motivates us to look for models with few K\"ahler moduli.

\subsec{Geometric considerations}

There are some general geometric constraints on the existence of
divisors of arithmetic genus $1$.  First, we can assume without loss
of generality that there exists a Weierstrass model
$\pi_0:W\rightarrow B$ and that $\mu:X\rightarrow W$ is the resolution
of the Weierstrass model. We recall that in F theory only the
vertical
divisors of arithmetic genus one
contribute to the superpotential.  As explained in \G,\ such divisors
fall in two classes: they are either components of the singular fibers
or they are of the form ${\cal V}=\pi^*(D)$ for $D$ a smooth divisor
on $B$.

To construct elliptic fourfolds which have divisors of the first type,
one can take $B$ to be a $\IP^1$-bundle over a surface $B^{\prime}$ and
enforce an ADE singularity of the Weierstrass model along
$B^{\prime}$, as in \katzvafa. The exceptional divisors of
$\mu:X\rightarrow W$ will have arithmetic genus one.  In \IIb\ string
theory terms, this corresponds to realizing pure Yang-Mills theory
with ADE gauge group on $7$-branes (D$7$-branes in the A and D cases,
and for the E case as in
\DasguptaIJ) and wrapping these branes on $B^\prime$.  The arithmetic
genus one condition then corresponds to the condition
$h^{0,1}(B^\prime)=h^{0,2}(B^\prime)=0$ for the resulting four
dimensional gauge theory not to have matter.
Because of the association of these divisors to brane gauge theory
we refer to them as ``gauge-type divisors.''

The other class of divisor, pullbacks of smooth divisors on $B$,
can give rise to D$3$-brane instanton corrections which need not have
a gauge theory interpretation, and so we refer to these as
``instanton-type divisors.''
In \refs{\W,\Mayr,\KLRY,\M} it was proposed, based on the study of
examples, that such divisors are always ``exceptional''
in the sense that there exist birational transformations of
Calabi-Yau fourfolds which contract these divisors. This was
shown to be true whenever $B$ is Fano in \G . Moreover,
when $B$ is Fano or toric, the number of the divisors contributing
to the superpotential is finite. This is because they are the
exceptional divisors associated to the contraction of one of the
generators of the Mori cone, which is polyhedral for $B$ Fano or
toric.\foot{There are other examples in which the number of contributing
divisors is infinite, such as the example of \DGW.}
These divisors have a negative intersection with the
corresponding generator of the Mori cone and are thus
``non-nef,'' as first pointed out in \DGW.

One can now see that there are no models with
$h^{1,1}=1$, because the negative intersection condition in this
case reduces to
$$
\Sigma \cdot D < 0
$$
where $D=\pi(\cV)$ and $\Sigma$ is an effective curve.\foot{
This inequality can be violated for gauge-type divisors, but we believe
all models containing these have $h^{1,1}>1$.
This point is discussed further in \RobbinsHX.}
The positivity
condition \kahlercone\ then forces the K\"ahler form to be $J=-t D$ with
$t>0$.  The conditions $V=J^3/6 > 0$ then forces
$Vol(D)=D\cdot J^2/2 < 0$,
so such an instanton correction cannot exist.  In words, the
``exceptional'' nature of these divisors means that in each of the
instanton amplitudes \npW, at least one of the coefficients in
the action $\vec n\cdot\vec \tau$ must be negative (in some basis).
Given several K\"ahler moduli and divisors, this need not be a
problem, and we will find in examples that it is not,
but it does preclude $h^{1,1}=1$.

A more mathematical argument for this point is that
if arithmetic genus one divisors are indeed always exceptional, then
there are no models with $h^{1,1}=1$, because contracting a divisor
will decrease $h^{1,1}$ (from the ``Contraction Theorem'' cited in
\G), which is impossible in this case.

It is also shown in \G\ that for $B$ Fano, divisors of arithmetic
genus one in fact have $h^{0,p}=0$ for $p>0$, so the necessary
condition for an instanton correction is in fact sufficient.

\newsec{Details of flux stabilization}

In the next section, we will find that models which will stabilize
all K\"ahler moduli have many complex structure moduli, $n\sim
100$. We will then need to find choices of flux which stabilize
the other moduli and lead to small $W_0$.  While one can search
for solutions numerically, as done in previous work
\refs{\kst,\GiryavetsVD,\MooreFG}, computation time becomes an
important issue. In particular the most straightforward approach
of picking arbitrary flux vectors, trying to find solutions to
$DW_0=0$, and hoping the solutions satisfy the desired properties,
especially that of having small but non-zero $W_0$,
becomes infeasible. One can simplify the problem somewhat by
imposing discrete symmetries, as we discuss below, but this of course
ignores most of the possible vacua.

A simpler goal is to obtain information about the existence and
number of flux vacua using the indirect approach of
\refs{\ad,\dd,\dsz}. We want to know how many supersymmetric vacua
we can expect with $L \leq L_*$, $e^{K_0} |W_0|^2 \leq \lambda_*$, the
dilaton $\tau$ within a region $\hat{\cal H} \subseteq {\cal H}$
and the complex structure $z$ within $\hat{\cal M}_c \subseteq
{\cal M}_c$, in the limit of very small $\lambda_*$. By
approximating the sum over flux lattice points by a volume
integral in continuous flux space, this was computed in \dd\ to be
 \eqn\Nflux{
 N_{vac} = {(2\pi L_*)^{b_3}\over b_3!} {\lambda_* \over L_*}
{\rm vol}(\hat{\cal H}) \int_{\hat{\cal M}_c}
 d^{2n} \! z \, \det g \,\, \rho_0(z)
 }
where $b_3=2n+2$, $g$ is the Weil-Petersson metric on ${\cal
M}_c$, and $\rho_0$ a certain density function on ${\cal M}_c$
computed from local geometric data. The detailed expression for
$\rho_0$ and a discussion of its evaluation for large $n$ is given
in appendix A. A useful estimate is the index density \ad:
 \eqn\index{
 \det g \, \rho_0 \sim {1 \over \pi^{n+1}} \det(R + \omega)
 }
with $R$ the curvature form and $\omega$ the K\"ahler form on
${\cal M}_c$. In particular $\rho_0 \neq 0$, so the distribution
of vacua with $|W_0|^2<\lambda_*$ is uniform in $\lambda_*$ around
$\lambda_*\sim 0$.

Performing the integral \Nflux\ over a region of moduli space
provides an estimate for the number of quantized flux vacua in
that region. While the estimate only becomes precise in the limit
of large $L_*$, numerical experiments suggest it is fairly
good for $L_* > b_3$. In this case, one expects a subregion of
radius $r > \sqrt{b_3/L_*}$ with an expected number of vacua $N_{vac} \gg 1$
to contain flux vacua \dd.

Once we know flux vacua exist in some region, there are better ways to
find explicit flux vacua.  We have developed a method which begins by
fixing a rational point at large complex structure, finding the
lattice of fluxes solving $DW=0$ and finding short lattice vectors
using advanced algorithms \cohen.  One can then move in by
systematically correcting the point on moduli space to take into
account the corrections from this limit.  Since these corrections are
small, this often produces vacua with small $W_0$, and we will cite
some results obtained this way below.

\subsec{Metastability}

Although we will not study the question of whether one can
break supersymmetry by antibranes or D terms in any detail, it leads
to another constraint on the flux vacua which we will study: namely,
we must insist that the potential is actually minimized at the
candidate vacuum.  This was not required for consistency of
a supersymmetric AdS vacuum \breit, and is trivial for a no-scale
nonsupersymmetric vacuum, but this condition becomes non-trivial
after size modulus stabilization.

As discussed in \dd\ and many other references, the mass matrix
for a vacuum satisfying $DW=0$, and for fields which do not
participate in the D-type supersymmetry breaking, is
 $$ m^2 = H^2 - 3 |\tilde{W}| H , $$
where we defined $\tilde{W} \equiv e^{K/2}|W|$, and the matrix
 $$ H = 2 \, d^2 |\tilde{W}|, $$
expressed in an orthonormal frame. Thus, positive eigenvalues of
$H$ which are less than $3|\tilde{W}|$ will lead to tachyons. In
the KKLT construction, $W_0$ is assumed small, and we show below
that this implies that $|W|$ is small, so that this need not be a
stringent condition.

We again take the superpotential to be a sum
\eqn\sumW{
 W = W_0(z,\tau) + W_\rho(\rho).
}
The matrix $H$ is given by
 $$
 H = |\tilde{W}| \cdot{\bf 1} +
 {1\over |\tilde{W}|}\left(\matrix{
 0& S& 0& T \cr
 \bar S& 0& \bar T& 0\cr
 0& T^t & 0& U \cr
 \bar{T}^t & 0& \bar U& 0}\right)
 $$
with
 $$
 S = {\bar{\tilde{W}}} D_iD_j \tilde{W}, \qquad
 U = {\bar{\tilde{W}}} D_\alpha D_\beta \tilde{W}, \qquad
 T = {\bar{\tilde{W}}} D_i D_\beta \tilde{W},
 $$
where $i,j$ are orthonormal frame indices for the complex
structure moduli and the dilaton, and $\alpha, \beta$ for the
K\"ahler moduli.\foot{
It may be counterintuitive that the mixing $T$ is present
as naively both $K$ and $W$ are the sum of two independent functions,
so the two sectors appear to decouple.
One way to see why this is naive is to note that this statement is not
invariant under K\"ahler-Weyl transformations, even those which take
the factorized form $W \rightarrow f_0(z,\tau)f_\rho(\rho)W$.
A more mathematical way to say this is that the decomposition
$K=K(z,\tau)+K(\rho)$ implies that $W$ is a section of a tensor product
line bundle, and in writing \sumW\ one is implicitly choosing reference
sections of the two bundles.  The mixing matrix $T$ is then
$-|W|^2$ times the tensor
product of the covariant derivatives of these reference sections.
}
If $S \gg U,T$, as is generically the case in the
KKLT construction, the effect of $T$ on the eigenvalues of $H$ will
be subleading. This expresses that when there is a large scale
separation of complex and K\"ahler moduli masses, they can be
treated separately. Moreover in general, if tachyons are found
when $T$ is set to zero (i.e.\ purely in complex structure or
K\"ahler directions), there will also be tachyons in the full
problem. Thus it is a good and useful approximation to put $T
\equiv 0$, and we will do so in the following.

The condition on the complex structure moduli is then the same as
in the discussion of \dd\ in which K\"ahler moduli were ignored,
except that $W$ is shifted to include $W_\rho$. This sets the
overall scale of the condition on $D^2W_{flux}$ but does not enter
in the details. For the simplest model superpotential
 $$ W_\rho = b \, e^{i\rho/N} , $$
we have
 $$ W = \left(1-{1\over 1+2 \, \Im \rho/3N}\right)W_0 $$
at $D_\rho W=0$, so for typical $\rho$ and $N$ this decreases $W$
and makes tachyons less likely, but not dramatically so.

For generic mass matrices, in a model with $n$ complex structure
moduli, some of these would be tachyonic in roughly a
fraction $3n|W|$ of cases, which is small for small $|W|$.

Going on to consider the K\"ahler moduli, for one K\"ahler
modulus, there are no tachyons if
 $$
 g^{\rho \bar{\rho}} |D_\rho^2W| > 2|W|
 $$
which is
 $$
 {4(\Im\rho)^2 \over 3} \left|\p_\rho^2 W_\rho
  - {3\over2 i\Im\rho}\p_\rho W_\rho - {3\over4(\Im\rho)^2}W\right|
  > 2|W| .
 $$
For the model superpotential this becomes
 $$
 \left| 1 + {\Im \rho \over N} \right| > 1
 $$
which is always satisfied. This might fail in a multi-modulus
model, though the examples we study below were found numerically
to be tachyon-free as well.

Thus, in general one does not expect moduli to become tachyonic after
supersymmetry breaking.  However special structure in the mass matrix
might change this conclusion, and the main point of this discussion is
that (in the approximation that we ignore the one-loop determinants in
$W_\rho$) the mass matrix for complex structure moduli can be analyzed
in the simpler model in which K\"ahler moduli are simply left out of
consideration, since their effect is just to renormalize $W_0$.
This justifies the analysis of \refs{\ad,\dd}\ in which this was done.

In \dd, the one parameter models were studied in great detail, and it
was found that tachyons are generic in some regimes, for example near
conifold points.  This is potentially important as we might want to
work near a conifold point to obtain a small scale or small
supersymmetry breaking.  The situation for multi-modulus models
appears to depend on details of the specific model, and this might or
might not be a problem.

\newsec{Search for models}

The upshot of the previous section is that a model which stabilizes
K\"ahler moduli must be based on a fourfold $X$ such that the divisors
on $Z$ whose pullbacks have arithmetic genus one form a ``complete
set'' as discussed in subsection 3.3.  The simplest way this can
happen is if such divisors form a basis of $H_4(Z)$.  We now show this
is a rather strong requirement and that not very many models satisfy
it.

There are two large classes of examples we consider: Fano threefolds and
$\IP^1$ bundles over toric surfaces.

First, since
there is a classification theory of Fano threefolds, all such examples
can be listed.  In particular, \G\ lists all the Fano threefolds
together with the divisors that give rise to divisors of arithmetic
genus one in the associated elliptically fibered fourfold.

Out of all the toric threefolds in the tables in \G, the models with a basis
of divisors of arithmetic genus $1$ are $B=\cF_{11}$, $\cF_{12}$, $\cF_{14}$,
$\cF_{15}$, $\cF_{16}$ and $\cF_{18}$.  The model with $B={\cF}_{17}$ comes
close,
but while it has enough fourfold divisors with arithmetic genus one,
the corresponding divisors in the base do not generate the Picard
group of ${\cal F}_{17}$.  Thus, out of the $18$ toric Fano manifolds in the
list, only six work. We note that there are also $74$ Fano threefolds that are
not toric. Out of these, only $23$ have enough fourfold divisors of arithmetic
genus one.

We consider $\cF_{18}$ and $\cF_{11}$ in more detail below, because
their orientifold limits have (relatively) few complex structure
parameters.

Another large class of examples can be obtained as $\IP^1$ bundles
over toric surfaces:
$$\xymatrix{
  \IP^1    \ar[r] &  B \ar[d]^{\pi'}\\
  & B'.\\
}$$
Let us start with the case  $B'=\IP^2$. The $\IP^1$-bundle over $B'$ will
be specified by an integer $n$ according to the following toric data:
\eqn\toricBii{
\matrix{ &  D_1 & D_2 & D_3 & D_4 & D_5 \cr
\IC^* & 1 & 1 & 1 & n & 0 \cr
\IC^* & 0 & 0 & 0 & 1 & 1.}}
$D_1,\ldots,D_5$ are the toric divisors : $D_1,D_2,D_3$ are the pullbacks
of the three lines in $\IP^2$ and $D_4$ and $D_5$ are the two
sections of the $\IP^1$ fibration. The arithmetic genus of
the fourfold divisors ${\cal V}_i=\pi^*(D_i)$ is given by $\chi({\cal
V}_i,{\cal O}_{{\cal V}_i})=1/2K_BD_i^2,i=1,\ldots,5$, where $K_B$ is
the canonical divisor of $B$ \G .  An immediate computation gives
\eqn\argeni{
\chi({\cal V}_1)=\chi({\cal V}_2)=\chi({\cal V}_3)=-1,~~\chi({\cal V}_4)=
-{n(n+3)\over 2},~~\chi({\cal V}_5)={n(-n+3)\over 2}.}
Now we see that that the pullbacks to the fourfold of the two sections
can not have simultaneously arithmetic genus $1$. However, let us first choose
$n$ to be $1$ or $-1$ so that either ${\cal V}_4$ or ${\cal V}_5$
will have arithmetic genus $1$. Now we can enforce an ADE type singularity along
the section whose pullback does not contribute initially to the superpotential.
The
components of the singular fiber will have arithmetic genus $1$ and will project
to that section. Since the two sections are linearly independent, they generate
the
Picard group of $B$ and we have thus obtained a model where the condition for
stabilizing all the K\"ahler moduli is satisfied.

Note that in this case it is possible to stabilize the K\"ahler moduli using
only
divisors of gauge type, by enforcing ADE
singularities of the Weierstrass model along both
of the two sections of the $\IP^1$ bundle. However,
this will not be true for models with $h^{1,1}>2$.

We can perform a similar analysis in the case when $B'$ is a Hirzebruch
surface $\IF_m$; The toric threefold $B$ is specified by two positive integers
$n$ and $p$ as $B=\IP({\cal O}_{\IF_m}\oplus{\cal O}_{\IF_m}(nC_0+pf))$, where
$C_0$ and $f$ are respectively the negative section and fiber of
$\IF_m$. The toric data of $B$ is
\eqn\toricBi{
\matrix{ &  D_1 & D_2 & D_3 & D_4 & D_5 & D_6\cr
\IC^* & 1 & 1 & m & 0 & p & 0 \cr
\IC^* & 0 & 0 & 1 & 1 & n & 0 \cr
\IC^* & 0 & 0 & 0 & 0 & 1 & 1.}}
$D_i, i=1,\ldots,6$ denote again the toric divisors. Another quick computation
gives
\eqn\argen{\eqalign{
&\chi({\cal V}_1)=0,~~\chi({\cal V}_2)=0,~~\chi({\cal V}_3)=-m,
~~\chi({\cal V}_4)=m,\cr &\chi({\cal V}_5)=-n-p-np+{mn(n+1)\over 2},
~~\chi({\cal V}_6)=n+p-np-{mn(-n+1)\over 2}.  }}
In order to satisfy
the requirement stated above we need to have $m=1$.  It is easy to see
that we can not have $\chi({\cal V}_5)=\chi({\cal V}_6)=1$.  These
equations would imply that $n$ and $p$ should satisfy $(n-p)^2-p^2=2$
which however does not admit integer solutions. But since $\chi({\cal
V}_5)$ and $\chi({\cal V}_6)$ are the the pullbacks of the two
sections we can enforce again an ADE singularity of the Weierstrass
model along one of them while choosing either $n=2(p+1)$ or $n=2(p-1)$
such that the pullback of the other section has also arithmetic genus
$1$. Obviously, in this case we can not construct a basis of the
Picard group of $B$ consisting only of arithmetic genus one divisors
of gauge type, because $h^{1,1} = 3$.

A similar analysis can be carried in the case when $B$ is a $\IP^1$
bundle over a toric del Pezzo surface. By making specific choices for
the data of the $\IP^1$ bundle, in the case of the del Pezzo surfaces
$dP_2,dP_3$ and $dP_4$, we can construct examples that have a full
basis of divisors of instanton type contributing to the
superpotential. What happens if we also consider divisors of gauge
type? For any $\IP^1$ bundle over $dP_2$ or $dP_3$, enforcing for
example an ADE singularity along one of the sections, we obtain a
model satisfying the above criterion. In the case of $dP_4$ we have to
enforce singularities of the Weierstrass model along both sections of
the $\IP^1$ bundle.

To summarize, there are several models with toric Fano base in which
instanton-type divisors can stabilize K\"ahler moduli.  In these
models, which can be analyzed using existing techniques, the presence
of a suitable nonperturbative superpotential is clear.

There are also several possibilities for models with $\IP^1$-fibered
base in which gauge-type divisors can stabilize K\"ahler moduli.
These models have heterotic duals and are potentially simpler, but
establishing the existence of a suitable superpotential in these
models requires controlling the matter content and matter
superpotential of the gauge theories.  This is a rather complicated
problem in the F theory framework, which has not been solved in the
detail we need; in particular the flux contributions to the matter
superpotential are not yet well understood.  Thus, we will not reach
definite conclusions for these models in this work.

Finally, there are surely many more models in which the base is not
Fano (any model with $h^{1,1}>10$ is necessarily of this type), and
there may also be models whose base has other fibration structures.

\subsec{The $\cF_{18}$ model}

One of the examples from \G\ is  $B\equiv{\cal F}_{18}$,
a toric Fano threefold \refs{\M,\MMi,\MMii}.  By our previous
discussion, $Z$ will be a double cover of $B$,
branched along its canonical divisor.
Thus $Z$ can be realized as a quadric in
$Y=\IP\big(\CO_{{\cal F}_{18}}\oplus\
\CO_{{\cal F}_{18}}(K_{{\cal F}_{18}})\big)$.

Note that $Y$ is not a weighted projective space and its toric data
is given by
\eqn\toricYi{
\matrix{ &  X_1 & X_2 & X_3 & X_4 & X_5 & X_6 & X_7 & X_8 & U & W \cr
\IC^* & -1 & 0 & 0 & 0 & 0 & 0 & 1 & 1 & -1 & 0 \cr
\IC^* & 1 & 1 & 0 & -1 & 0 & 0 & 0 & 0 & -1 & 0 \cr
\IC^* & 0 & 1 & 1 & 0 & -1 & 0 & 0 & 0 & -1 & 0 \cr
\IC^* & 1 & 0 & 1 & 0 & 0 & -1 & 0 & 0 & -1 & 0 \cr
\IC^* & 0 & -1 & 0 & 1 & 1 & 0 & 0 & 0 & -1 & 0 \cr
\IC^* & -1 & 0 & 0 & 1 & 0 & 1 & 0 & 0 & -1 & 0 \cr
\IC^* & 0 & 0 & -1 & 0 & 1 & 1 & 0 & 0 & -1 & 0 \cr
\IC^* & 0 & 0 & 0 & 0 & 0 & 0 & 0 & 0 & 1 & 1.}}

The generic divisor in the linear system $|-K_Y|$ is the smooth
Calabi-Yau variety $Z$ and is defined by the equation
\eqn\eqZ{
W^2+WU\sum_{(a_1,\ldots, a_8)'}f_{a_1\ldots a_8}\prod_{i=1}^8 X_i^{a_i}+
U^2\sum_{(b_1,\ldots, b_8)'}g_{b_1\ldots b_8}\prod_{i=1}^8 X_i^{b_i}=0,
}
where the sums are taken over sets of positive integers
$a_1,\ldots,a_8,b_1,\ldots,b_8$ which satisfy
\eqn\setcons{\vbox{\halign{ $#$ \hfill &\qquad  $#$ \hfill\cr
-b_1+b_7+b_8=2(-a_1+a_7+a_8)=2, & b_1+b_2-b_4=2(a_1+a_2-a_4)=2,\cr
-b_1+b_4+b_6=2(-a_1+a_4+a_6)=2, & b_1+b_3-b_6=2(a_1+a_3-a_6)=2,\cr
-b_2+b_4+b_5=2(-a_2+a_4+a_5)=2, & b_2+b_3-b_5=2(a_2+a_3-a_5)=2,\cr
-b_3+b_5+b_6=2(-a_3+a_5+a_6)=2. & \cr
}}}
Using the invariance under reparametrizations,
we can set $f_{a_1,\ldots,a_8}=0$ and the hypersurface equation becomes
\eqn\eqZi{
W^2+U^2\sum_{(b_1,\ldots, b_8)'}g_{b_1\ldots b_8}\prod_{i=1}^8 X_i^{b_i}=0.
}

The holomorphic involution $\bar \Omega$ is simply $W \rightarrow
-W$. All the third cohomology of $Z$ is odd under the (pull-back)
of $\hat\Omega$ and in particular $\hat\Omega ^*\Omega=-\Omega$,
where $\Omega$ is the holomorphic three-form on $Z$.  The third
cohomology of the quotient $B=Z/\hat\Omega$ is trivial since $B$
is toric. Conversely, $H^2(Z,\IZ)$ and $H^4(Z,\IZ)$ are even under
$\hat\Omega^*$. Therefore, all the complex structure and K\"ahler
deformations remain in the spectrum \BH .

Thus, we
obtain a IIB orientifold compactification on $B$ of the type we want.

\subsubsec{{Topological analysis}}

This can be done using standard toric techniques \B .
The toric data of ${\cal F}_{18}$ is \refs{\KLRY,\M,\MMi,\MMii}
\eqn\toricBi{
\matrix{ & D_1 & D_2 & D_3 & D_4 & D_5 & D_6 & D_7 & D_8\cr
\IC^* & -1 & 0 & 0 & 0 & 0 & 0 & 1 & 1 \cr
\IC^* & 1 & 1 & 0 & -1 & 0 & 0 & 0 & 0 \cr
\IC^* & 0 & 1 & 1 & 0 & -1 & 0 & 0 & 0 \cr
\IC^* & 1 & 0 & 1 & 0 & 0 & -1 & 0 & 0 \cr
\IC^* & 0 & -1 & 0 & 1 & 1 & 0 & 0 & 0 \cr
\IC^* & -1 & 0 & 0 & 1 & 0 & 1 & 0 & 0 \cr
\IC^* & 0 & 0 & -1 & 0 & 1 & 1 & 0 & 0.}}

We choose $D_i,i=1,\dots,5$ as basis for $H^2(B,\IZ)$.
We have the linear equivalence relations
$D_6=D_2-D_3+D_4,D_7=D_8=-D_1+D_3-D_4+D_5$.

The K\"ahler cone is not simplicial: although it is five dimensional,
it has six generators which are given by
\eqn\Kgeni{\eqalign{
&{\cal R}_1=D_7,~~{\cal R}_2=D_1+D_6+D_7,~~{\cal
R}_3=D_1+D_3+D_6+D_7,~~{\cal R}_4=D_3+D_6,\cr &{\cal
R}_5=D_1-D_2+D_3+D_6+D_7,~~{\cal R}_6=D_1-D_2+D_3+2D_6+D_7.  }} In
order to define a large radius limit and K\"ahler coordinates, we need
to choose a simplicial decomposition of the K\"ahler cone and pick one
of the subcones \CKYZ . We note that there exists a simplicial
decomposition such that one of the subcones is generated by ${\cal
R}_j$, $j=1,\ldots,5$ and therefore we take the K\"ahler form to be
\eqn\Kahleri{\eqalign{
J=\sum_{j=1}^5t_j
 {\cal R}_j&=(t_2+t_3+t_5)D_1-t_5D_2+(t_3+t_4+t_5)D_3+(t_2+t_3+t_4+t_5)D_6\cr
&+(t_1+t_2+t_3+t_5)D_7.}}

The dual polyhedron
$\nabla_Z$ which encodes the
divisors of the Calabi-Yau threefold $Z$ has the following vertices:
\eqn\verti{\matrix{
{\tt (0, 0, -1, 1)}, & {\tt (0, 1, 1, 1)}, & {\tt (0, -1, 0, 1)}, &
{\tt (0, 1, 0, 1)}, & {\tt (0, 0, 1, 1),}\cr
\cr
{\tt (0, -1, -1, 1)}, & {\tt (1, 0, 0, 1)}, & {\tt (-1, 0, -1, 1)}, &
{\tt (0, 0, 0, -1).}\cr}}
The Hodge numbers\foot{We have used the computer program {\tt
POLYHEDRON}, written by Philip Candelas.} of
$Z$ are $h^{1,1}(Z)=5,h^{2,1}(Z)=89$.

Next, we need to know how divisors of $B$ pull back to the
Calabi-Yau fourfold $X$.
Since $X$ elliptically fibered over the base $B={\cal F}_{18}$,
we can construct it as a hypersurface in a toric variety.

Using this, we can construct the dual polyhedron
$\nabla_X$, which encodes the divisors of the Calabi-Yau fourfold $X$.
It has vertices:
\eqn\verti{
\matrix{{\tt (0, 0, -1, 2, 3)}, & {\tt (0, 1, 1, 2, 3)},
& {\tt (0, -1, 0, 2, 3)}, & {\tt (0, 1, 0, 2, 3)}, & {\tt (0, 0, 1, 2, 3)},\cr
\cr
{\tt (0, -1, -1, 2, 3)}, & {\tt (1, 0, 0, 2, 3)}, & {\tt (-1, 0, -1,
2, 3)}, & {\tt (0, 0, 0, -1, 0)}, & {\tt (0, 0, 0, 0, -1).}\cr}}
Standard toric methods give for the Hodge numbers
of the fourfold
$h^{1,1}=6,~h^{2,1}=0,~h^{3,1}=2194,~h^{2,2}=8844,~\chi=13248$. We
note that it is possible to find a triangulation of $\nabla_X$
consistent with its elliptic fibration structure such that each of the
top dimensional cones having unit volume, thus guaranteeing smoothness
of the corresponding Calabi-Yau fourfold.

\subsubsec{Nonperturbative superpotential}

The fibration $\pi:X\longrightarrow B$ has a section
$\sigma:B\longrightarrow X$, $\pi\circ\sigma=\bbbone_B$. The following
toric divisors have arithmetic genus one: ${\cal V}_1={\tt(0, 0, -1, 2,
3)},~{\cal V}_2={\tt(0, 1, 1, 2, 3)}, ~{\cal V}_3={\tt(0, -1, 0, 2,
3)},~{\cal V}_4={\tt(0, 1, 0, 2, 3)},~{\cal V}_5={\tt(0, 0, 1, 2,
3)},~{\cal V}_6= {\tt(0, -1, -1, 2, 3)}$ and $\Sigma={\tt (0, 0, 0, 2,
3)}$. The first six divisors are vertical, ${\cal V}_i=\pi^*(D_i)$,
$i=1,\ldots,6$, and therefore contribute to the F-theory
superpotential, while $\Sigma$ does not contribute, since is the
section of the elliptic fibration, $\Sigma=\sigma(B)$. Moreover, since
${\cal V}_j,j=1,\ldots,6$ are vertices of $\nabla_X$, $h^{0,i}({\cal
V}_j)=0,i=1,2,3,j=1,\ldots,6$. Now, since $D_1,\ldots,D_5$ generate
$H^2(B,\IZ)$, we see that this model has a complete set
of contributing divisors and satisfies the requirement of
section 3.

Note that the divisor $D_1$ on the base $B$ is the exceptional divisor
that corresponds to the contractions of both first and sixth Mori cone
generators.
This is possible since $D_1$ is isomorphic to a Hirzebruch surface $\IF_0$
which is a product of those two curves.

In order to study the question of fixing the K\"ahler moduli, we need
to compute the volumes of the divisors contributing to the
superpotential, as well the volume of the three dimensional base
$B$. To achieve that, we triangulate the fan of $B$. We list the cones
below:
\eqn\conesi{\eqalign{
&D_1D_4D_7,~~D_1D_4D_8,~~D_1D_6D_7,~~D_1D_6D_8,~~D_2D_4D_7,~~D_2D_4D_8,\cr
&D_2D_5D_7,~~D_2D_5D_8,~~D_3D_5D_7,~~D_3D_5D_8,~~D_3D_6D_7,~~D_3D_6D_8.}}

Using the above (unique) triangulation, we obtain the following nonvanishing
intersection numbers
\eqn\setcons{\vbox{\halign{ $#$ \hfill &\qquad  $#$ \hfill
&\qquad  $#$ \hfill  &\qquad  $#$ \hfill &\qquad  $#$ \hfill\cr
D_1^3=2, & D_1^2D_4=-1, & D_1^2D_6=-1, & D_1^2D_7=-1, & D_2^3=-1,\cr
D_2^2D_4=1, & D_2^2D_7=-1, & D_3^3=-1, & D_3^2D_6=1, & D_3^2D_7=-1,\cr
D_4^2D_2=-1, & D_4^3=1, & D_4^2D_7=-1, & D_5^2D_2=1, & D_5^2D_3=1,\cr
D_5^3=-2, & D_5^2D_7=-1, & D_6^2D_3=-1, & D_6^3=1, & D_6^2D_7=-1.\cr
}}}
 We obtain for the divisor volumes $\tau_i
\equiv D_i J^2/2$ and the total volume $V \equiv J^3/6$:
\eqn\dvols{\eqalign{
 &\tau_1 = t_1t_4,~~\tau_2 = {t_5\over 2}(2t_1+2t_2+2t_3+t_5), \cr
 &\tau_3 = {t_2\over 2}(2t_1+t_2+2t_3+2t_5),~~\tau_4={1\over
2}(t_2+t_3)(2t_1+t_2+t_3) \cr
 &\tau_5 = (t_3+t_4)(t_1+t_2+t_3+t_5),~~\tau_6={1\over
2}(t_3+t_5)(2t_1+t_3+t_5), \cr
 &V = t_1t_2t_3 + {t_2^2t_3\over 2} + {t_1t_3^2\over 2} + t_2t_3^2 + {t_3^3\over
3} +
    t_1t_2t_4 + {t_2^2t_4\over 2} + t_1t_3t_4 +
    t_2t_3t_4 + {t_3^2t_4\over 2}\cr
&\qquad+ t_1t_2t_5 + {t_2^2t_5\over 2} +
    t_1t_3t_5 + 2t_2t_3t_5 + t_3^2t_5 + t_1t_4t_5 + t_2t_4t_5 +
    t_3t_4t_5 + {t_2t_5^2\over 2}\cr
&\qquad+ {t_3t_5^2\over 2} + {t_4t_5^2\over 2}.
}}

\subsubsec{{Complex Structure Moduli}}

To compute or count flux vacua in arbitrary regions of the complex
structure moduli space of the Calabi-Yau threefold $Z$, one would
have to compute the periods for the generic hypersurface $Z$.
These are generalized hypergeometric functions in 89 variables. In
principle they can be computed using existing techniques, but this
would require a lot of work, even using a computer.

A somewhat easier (but still formidable) task is
to describe the periods in the vicinity of the large complex
structure limit. This is equivalent to computing the triple
intersections for the mirror threefold ${\tilde Z}$, which has
$h^{2,1}=5$ and $h^{1,1}=89$. To make the description of this part
of the moduli space complete, one furthermore has to compute the
K\"ahler cone, i.e.\ the part of the parameter space $\IR^{89}$
where all holomorphic curves have positive area. We describe the
algorithms we used to achieve this in appendix B.

\subsubsec{Flux vacua}

Counting flux vacua can be done using the techniques of
\refs{\ad,\dd} as outlined in section 4. Since $L_*=13248/24=552
\gg b_3=180$, we expect that the approximations made to derive the
counting formula \Nflux\ should be valid.

According to this formula, the expected number of vacua with
cosmological constant $e^{K_0} |W_0|^2$ less than $\lambda_* \ll L_*$,
equals $(2\pi L)^{b_3}/b_3! \sim 10^{307}$ multiplied by
$\lambda_*/L_*$ times the integral of a geometrical density
function. Taking the integration domain $\hat{{\cal M}}_c$ equal
to the entire moduli space, the estimate \index\ indicates that
the geometrical factor should be of the order of the Euler
characteristic of ${\cal M}_c$. This does not need to be an
integer, as the moduli space is a noncompact orbifold, and in
examples \dd\ is a small fraction, of order
$1/|\Gamma|$, where $\Gamma$ is the order of a finite group
(e.g.\ $\IZ_5$ for the mirror quintic) or the volume of a group \lu.
This will be far larger than $10^{-307}$ and
thus we expect many vacua with very small
cosmological constant.

Can we find such vacua at large complex structure, or
equivalently, at large volume of the mirror? We can judge
this by integrating the density function over, say,
the region defined by requiring all curve areas of the mirror to
be bigger than 1. Using the approach for estimating the density
function $\rho_0$ explained in appendix A and the construction of
the geometry of the moduli space outlined in appendix B, we have
done Monte Carlo estimates of this integral. The results are as
follows. The average value of $\mu$ defined in appendix A is of
order $10^{90}$, hence the density function $\rho_0$ is of order
$10^{240}$, and we take this out of the integral \Nflux. The
remaining volume integral, evaluated using $10^7$ Monte Carlo
sample points, gives a number of order $10^{-650}$. Putting
everything together we get, up to the factor $\lambda_*/L_*$,
 \eqn\Nvacresult{
 N_{vac}(LCS) \sim 10^{-100}.
 }
Why is the volume so small? One important reason is the mirror
volume suppression factor $\tV^{-n/3}$ mentioned at the end of
appendix A. For curve areas $y^i$ bigger than one, the mirror
volume $\tV > 10^{10}$. This becomes understandable when one
considers that when $\tV$ is written in terms of the curve area
coordinates $y^i$ (cf.\ appendix A), the expression $\tV =
\tD^{(y)}_{ijk} y^i y^j y^k/6$ contains $\sim 10^6$ terms, and the
$\tD^{(y)}_{ijk}$ are widely distributed between 0 and $10^8$.
With $\tV > 10^{10}$, the volume suppression factor is $\sim
10^{-300}$. Additional suppression comes from the fact that $\det
g \sim 10^{-100}$ for typical values of $y$ with $\tV(y)=1$, and
from the smallness of the Euclidean volume of the surface
$\tV(y)=1$. It is possible that the Monte Carlo missed regions in
which $\tV$ is smaller than $10^{10}$, or that some of this is an
artifact of our approximate parametrization of the K\"ahler cone
with the curve areas $y^i$, but we see no particular evidence for
this.

Thus it seems likely that any flux vacua in this region are
special cases, and there is no reason to expect a multiplicity of
vacua out of which some would have small $W_0$. This paucity of
vacua in the large complex structure limit is not specific to this
example, but rather is a very general feature of models with many
moduli, as explained in \ddtwo.

If we neglect world-sheet instanton corrections, then by lowering
the cutoff to $y^i > 0.075$, the expected number of vacua becomes of order 1.
In few modulus examples, the instanton sums tend to converge all the
way down to zero, so this might be a valid indication of where vacua will
start to exist.

\subsec{The $\cF_{11}$ model.}

Another example from Grassi's list is the base $B={\cal F}_{11}$
\refs{\MMi,\MMii}.
We now construct an explicit toric model of a Calabi-Yau fourfold
elliptically fibered over this base. This model has an orientifold
limit $Z$ with $h^{1,1}=3,h^{2,1}=111$.

The toric data of ${\cal F}_{11}$ is given by
\refs{\KLRY,\M,\MMi,\MMii}
\eqn\toricB{
\matrix{ & D_1 & D_2 & D_3 & D_4 & D_5 & D_6 \cr
\IC^* & 0 & -2 & 1 & 1 & 1 & 0  \cr
\IC^* & -1 & 0 & 1 & 0 & 0 & 1  \cr
\IC^* & 1 & 1 & -1 & 0 & 0 & 0. \cr}}
The generators of the K\"ahler cone of $B$ are
\eqn\genK{
{\cal R}_1=D_5,~~{\cal R}_2=D_6,~~{\cal R}_3=D_1+D_6.
}

We have the linear equivalence relations $D_4=D_5=D_1+D_3$ and
$D_6=D_1+D_2+2D_3$. The Calabi-Yau fourfold elliptically fibered over
$B$ may be constructed as a hypersurface in a toric variety. The dual
polyhedron $\nabla$, which encode the divisors of the Calabi-Yau
fourfold has vertices:
$$\eqalign{&{\tt (1, 1, 0, 2, 3),~~(0, -1, 0, 2, 3),~~(1, 0, 0, 2,
3),~~(0, -1, 1, 2, 3),~~(-1, -1, -1, 2, 3),}\cr &{\tt (0, 1, 0, 2,
3),~~(0, 0, 0, -1, 0),~~(0, 0, 0, 0, -1).}\cr}$$ The Hodge numbers of
this fourfold are
$h^{1,1}=4,~h^{2,1}=0,~h^{3,1}=3036,~h^{2,2}=12204,~\chi=18288$. Again,
we find that the fourfold is smooth.

The fibration $\pi:X\longrightarrow B$ has a section
$\sigma:B\longrightarrow X$, $\pi\circ\sigma=\bbbone_B$. The following
divisors have arithmetic genus one: ${\cal V}_1={\tt(1, 1, 0, 2,
3)},~{\cal V}_2={\tt(0, -1, 0, 2, 3)}, ~{\cal V}_3={\tt(1, 0, 0, 2,
3)}$ and $\Sigma={\tt (0, 0, 0, 2, 3)}$. The first three divisors are
vertical, ${\cal V}_i=\pi^*(D_i)$, $i=1,2,3$, and therefore may
contribute to the F-theory superpotential, while $\Sigma$ does not
contribute, since is the section of the elliptic fibration,
$\Sigma=\sigma(B)$. It is possible to check that in fact
$h^{0,i}({\cal V}_j)=0, i,j=1,2,3$, thus all the vertical divisors
do give a contribution to the superpotential. Therefore, this model
also provides a complete basis of divisors.

To compute the volumes of the divisors $D_1,D_2,D_3$, we triangulate
the fan of $B$.  Its 3-dimensional cones are
$D_1D_4D_6,~D_1D_5D_6,~D_4D_5D_6,~D_2D_4D_5,~D_1D_3D_4,~D_1D_3D_5,$
$D_2D_3D_4,~D_2D_3D_5$. Using this (unique)
triangulation, we obtain the following nonvanishing triple
intersections:
\eqn\ti{
D_1^3=-3,~~D_1^2D_3=2,~~D_1 D_3^2=-1,~~D_2^3=4,~~D_2^2 D_3= -2,~~D_2 D_3^2 = 1.
}

Let $J = \sum_{i=1}^3 t_i {\cal R}_i$ be the K\"ahler form of $B$. We
obtain for the divisor volumes $\tau_i
\equiv D_i J^2/2$ and the total volume $V \equiv J^3/6$:
\eqn\dvols{\eqalign{
 &\tau_1 = {t_2\over 2}(2t_1+t_2+4t_3), ~~\tau_2 = {t_1^2\over
 2},~~\tau_3 = t_3(t_1+t_3), \cr &V = {t_1^2t_2\over 2}+{t_1t_2^2\over
 2}+{t_2^3\over 6}+{t_1^2t_3\over
 2}+2t_1t_2t_3+t_2^2t_3+t_1t_3^2+2t_2t_3^2+{2t_3^3\over 3}.  }}

\subsec{The orientifold of $\IP^4_{[1,1,1,6,9]}$}

Finally, out of the various possibilities which can be obtained as
$\IP^1$ bundles over a toric surface, the simplest is perhaps the
$\IP^1$ bundle over $\IP^2$. The toric data of the threefold is
presented in \toricBii,\ where we take $n=-6$, so that $B=\IP({\cal O}_{\IP^2}
\oplus{\cal O}_{\IP^2}(-6))$. The toric data for $B$ is as follows:
\eqn\toricBv{
\matrix{ &  D_1 & D_2 & D_3 & D_4 & D_5 \cr
\IC^* & 1 & 1 & 1 & -6 & 0 \cr
\IC^* & 0 & 0 & 0 & 1 & 1.\cr
}}
The toric divisors $D_4$ and $D_5$ are the sections of the $\IP^1$
bundle over $B'=\IP^2$. We will construct an elliptic fourfold $X$
over $B$ in such a way that $X$ will be the resolution of a
Weierstrass model $W$ which has a $D_4$ singularity along the first
section and an $E_6$ singularity along the second section. The
motivation for this choice will become clear later, when we study
K\"ahler moduli stabilization for these models


The dual polyhedron $\nabla_X$, which encode the divisors of the
Calabi -Yau fourfold has vertices:
$$
\eqalign{&{\tt (-1, -1, 6, 2, 3),~~(0, 0, -3, 2, 3),~~(0, 1, 0, 2,
3),~~(1, 0, 0, 2, 3),~~{(0,  0, -2, 1, 1),}}\cr &{\tt (0, 0, -1, 0, 0),~~
(0, 0, 0, -1, 0),~~(0, 0, 0, 0, -1).}\cr}
$$
The Hodge numbers of this fourfold are
$h^{1,1}=13,~h^{2,1}=0,~h^{3,1}=1071,~h^{2,2}=4380,~\chi=6552$. Again,
we find that the fourfold is smooth. We can also enforce an $E_7$ or
$E_8$ singularity along the infinity section. The data for the
fourfolds obtained this way are, respectively,
$h^{1,1}=14,~h^{2,1}=0,~h^{3,1}=935,~h^{2,2}=3840, ~\chi=5742$ and
$h^{1,1}=16,~h^{2,1}=253,~h^{3,1}=745,~h^{2,2}=2582,~\chi=3096$.

It is easy to check that in this case the orientifold limit $Z$
will be an elliptic fibration over $\IP^2$, which is familiar as the
hypersurface in weighted projective space $\IP^4_{[1,1,1,6,9]}$
studied in \refs{\candtwo,\drom} and several other works. In order to do that,
we note that $Z$ is given by a quadric in a toric variety $Y$ described by
the following data
\eqn\toricYi{
\matrix{ &  X_1 & X_2 & X_3 & X_4 & X_5 & U & W \cr
\IC^* & 1 & 1 & 1 & -6 & 0 & 3 & 0  \cr
\IC^* & 0 & 0 & 0 & 1 & 1 & -2 & 0  \cr
\IC^* & 0 & 0 & 0 & 0 & 0 & 1 & 1. \cr
}}
This threefold has $h^{1,1}=2$ and $h^{2,1}=272$.  There is a unique toric
Calabi-Yau threefold with this Hodge numbers, the elliptic fibration over
$\IP^2$ \KSi .
\medskip

{\it Nonperturbative superpotential}

We proceed as before and start by listing all toric divisors
of arithmetic genus one for the first of the models above. These are:
${\cal V}_1={\tt(0, 0, -3, 2,
3)},~{\cal V}_2={\tt(0, 0, -2, 1, 1)}, ~{\cal V}_3={\tt(0, 0, -2, 1,
2)},~{\cal V}_4={\tt(0, 0, -2, 2, 3)},~{\cal V}_5={\tt(0, 0, -1, 0,
0)}, ~{\cal V}_6={\tt(0, 0, -1, 0, 1)}, ~{\cal V}_7={\tt(0, 0, -1, 2,
3)},~{\cal V}_8={\tt(0, 0, 2, 2, 3)}, ~{\cal V}_9={\tt(0, 0, 1, 2, 3)}.$

The first seven divisors project to $D_5$ and the last two project to
$D_4$. These are going to be the only base divisors contributing to
the superpotential for the other two models as well.

To compute the volumes of the divisors $D_4$ and $D_5$ we start by
triangulating the fan of $B$. Its three dimensional cones are
$D_1D_2D_4,~D_1D_2D_5,~D_1D_3D_4,~D_1D_3D_5, ~D_2D_3D_4$ and
$D_2D_3D_5$. Using this triangulation we obtain the following nonzero
intersection numbers
\eqn\wpints{\eqalign{
&
D_1^2D_4=1,~~D_1^2D_5=1,~~D_2^2D_4=1,~~D_2^2D_5=1,~~D_3^2D_4=1,~~D_3^2D_5=1,\cr
& D_1D_4^2=-6,~~D_1D_5^2=6,~~D_2D_4^2=-6,~~D_2D_5^2=6,~~D_3D_4^2=-
6,~~D_3D_5^2=6,\cr
& D_4^3=36,~~ D_5^3=36.}}

Let $J=t_1D_1+t_5D_5$ be the K\"ahler form of $B$. We obtain for the
divisor volumes $\tau_i\equiv D_iJ^2/2$ and the total volume $V\equiv
J^3/6$:
 \eqn\volswp{\eqalign{
 &\tau_4={t_1^2\over 2},~~\tau_5={(t_1+6 t_5)^2\over 2},\cr
 & V={1\over 6}(3t_1^2t_5+18t_1t_5^2+36t_5^3).}}

\medskip
{\it Complex structure moduli and flux vacua}

The full 272 parameter prepotential for this model
has not been worked out.
However, this
moduli space admits a $\Gamma\equiv\IZ_6\times \IZ_{18}$ action,
which fixes the two parameter subspace of CY's with
defining equation
$$
f = x_1^{18}+x_2^{18}+x_3^{18}+x_4^3+x_5^2
 - 18\psi x_1x_2x_3x_4x_5 - 3\phi x_1^6x_2^6x_3^6 .
$$
This is the subset of CY's obtained by the mirror construction
of \GreeneUD\ and the six periods of these mirror CY's,
which is the same as the subset of periods of $\Gamma$-invariant
cycles, are worked out
in \candtwo.

If one turns on flux only on these cycles, since the resulting
superpotential and K\"ahler potential are $\Gamma$-invariant, one
is guaranteed that $D_iW=0$ in all $\Gamma$-noninvariant directions,
and thus one can find flux vacua just by working in the $\Gamma$-invariant
part of the moduli space, call this $\cM_\Gamma$.
(This observation is also made in \GiryavetsVD.)
Of course, one would eventually need to check that all moduli remain
non-tachyonic after supersymmetry breaking.

For the choices of gauge groups we considered,
$L=\chi/24 \sim 100 - 300$, so the total index for flux vacua in this subspace
is
\refs{\ad,\dd}
$$
I = {(2\pi L)^6\over 6!}
\int_{\cM_\Gamma\times\CH} \det(-R-\omega) .
$$
As mentioned earlier,
the integral can be estimated as $1/12|\Gamma|=1/1296$, leading to
$$
I \sim 10^{11} - 10^{13} .
$$
The number at weak coupling $1/g_s^2<\epsilon$ will be roughly
$\epsilon$ times this, and the number with $e^{K_0}|W_0|^2 < \lambda L T_3$
will be roughly $\lambda$ times this,
so there are clearly many weakly coupled flux
vacua with small $W_0$.

Just to get a few explicit flux vacua, we consider the region
of large complex structure,
{\it i.e.} the region in which instanton corrections are small,
$N_i e^{2\pi i w_i} << 1$.  From \candtwo, the prepotential
at third order in the world-sheet instanton expansion is
\eqn\instexp{\eqalign{\cF =
 &{1\over 6}\left(9 w_1^3 + 9 w_1^2 w_2 + 3 w_1 w_2^2\right) \cr
 &
 -{9\over 4}w_1^2 - {3 \over 2} w_1 w_2 - {17\over 4}w_1 - {3\over 2}w_2
 + \xi \cr
 &+ {1\over (2\pi i)^3}
   \bigg( 540 q_1+3q_2+{1215\over 2}q_1^2-1080q_1q_2 - {45\over 2}q_2^2 \cr
 &\qquad\qquad
  + 560 q_1^3 + 143370 q_1^2 q_2 + 2700 q_1 q_2^2 + {244\over 9}q_2^3
 +\ldots\bigg)
 }}
with $q_i\equiv e^{2\pi i w_i}$ and $\xi \equiv {\zeta(3) \chi(Z)
\over 2 (2 \pi i)^3} \approx -1.30843 \, i$.

Looking at the coefficients in this expansion, the instanton
corrections should be small for $w_2 >> 1/6$ and $w_1 > 1$.  To
find quantized flux vacua, we used the procedure discussed in
section 4. As an example, we first look for fluxes which stabilize
the moduli at the rational point $\tau=3i, w_1=i, w_2=i$, ignoring
world-sheet instanton corrections and rationally approximating
$\xi$ by $-13/10$. One finds a quantized flux vacuum with $W=0$,
with fluxes $(N_{RR};N_{NS})$ equal to
 $$
 \{ 0,69,28,0,0,-20;-49,-18,-6,-4,0,0 \};
 \qquad L=352
 $$
granting that the quantization condition on the orientifold is the
same as on the original CY. This was argued in general in \frey;
the fact that some cycles are smaller on the orientifold which
naively doubles the Dirac quantum, is compensated for by the
possibility of discrete RR and NS flux at the orientifold fixed
points.  A careful discussion in the present example might be
possible using K theory \mooreK.

Restoring the exact value of $\xi$, including the first instanton
corrections and solving the resulting equations $DW=0$ produces a
$W \neq 0$ vacuum near the starting point. We find
 $$\matrix{
 \tau&=&  2.945 i \cr
 w_1&=& 0.9625  i \cr
 w_2&=& 1.1037 i \cr
 e^K|W|^2&=& 1.379 \times 10^{-4}
 } $$
Examples with larger $\tau,w_i$ can easily be found in this way,
but $L$ tends to become bigger then as well.

For these vacua, one can check that the instanton corrections are
small (as one would guess by the small corrections they lead to
for the moduli) and the vacua appear sound, on the level we are
working.

\newsec{Numerical results on K\"ahler stabilization}

\subsec{${\cal F}_{18}$ model}

The K\"ahler moduli are stabilized for generic order 1 values of
the $b_i$. Taking $W_0 = 10^{-30}$, the typical values for the
$t_i$ are $t_1 \sim 50-100$, $t_i \sim 0.1 - 0.3$ for $i=2,4,5$
and $t_3 \sim 0$. The corresponding values of the contributing
divisor volumes are $V_i \sim 11-12$, and the total volume $V \sim
5-10$. Taking the $b_i$ all equal gives a nongeneric singular
solution: $t_1 = \infty$, $t_i = 0$ for $i>0$. But with
$b_i=(1,1.5,2,2.5,3,3.5)$ for example, we get
$t_i=(53.3,0.209,0.00156,0.208,0.222)$, $\tau_i = (11.1,11.9,11.2+
i \pi,11.2,11.3 + i\pi,11.9 + i\pi)$, and $V = 7.32$. Different
choices for the phases of $W_0$ and five of the $b_i$ can be
absorbed in shifts of the imaginary parts of the $\tau_i$ (the
axions). Note that since the areas of the Mori cone generators are
$(t_1,t_2+t_3,t_4+t_3,t_5+t_3,t_5,t_4,t_2)$, these solutions lie
well inside the K\"ahler cone; $t_3=0$ is only an interior
boundary of a subsimplex of the K\"ahler cone.

The typical values for the $t_i$ and the total volume are rather
small for this model, so $\alpha'$ corrections could become
important. (The situation is better for the other models we
consider.) However, since the solution in terms of the $\tau_i$ is
mainly determined by the exponential factors (hence the near-equal
values of $\tau_i \sim - (\ln W_0)/2 \pi \approx 11$), it is
reasonable to believe that it will still exist at approximately
the same values of $\tau_i$ even after taking into account such
corrections.

Finally, to have a stable vacuum after lifting the cosmological
constant to a positive value, the critical point has to be a
minimum of the potential. We verified numerically that this is
indeed the case.

\subsec{${\cal F}_{11}$ model}

The K\"ahler moduli are stabilized for generic order 1 values of
the $b_i$. For $W_0 = - 10^{-30}$, $b_i = 1$, we get
$t_i=(4.89,1.30,1.76)$. These are also the areas of the three
generators of the Mori cone. The corresponding complexified
volumes of the divisors $D_i$ contributing to the superpotential
are $\tau_i = (11.8,11.9,11.7)$ and the total volume is $V=93.3$.
The critical point is a minimum of the potential.

\subsec{$\IP^4_{[1,1,1,6,9]}$ model}

Assuming the gauge theory generates a superpotential $W=W_0 +
\sum_{i=1}^2 b_i e^{-2 \pi a_i \tau_i}$, where $\tau_1$, $\tau_2$
are the complexified volumes of the divisors $D_4$ and $D_5$, we
find that the K\"ahler moduli are stabilized for generic order 1
values of the $b_i$, provided $a_1 > a_2$. The following are some
of the values of $t_i$, $V_i$ and the volume $V$, obtained for
$b_i=1$ and different choices of $a_1$, $a_2$ and $W_0$:
 \eqn\numresults{
 \matrix{ a_1 & a_2 & W_0 & t_1 & t_5 & V_1 & V_2 & V  \cr
              &     &     &     &     &     &     &   & \cr
          1/4 & 1/30 & 10^{-30} & 9.83 & 2.76 & 48.3 & 348 & 484 \cr
          1/4 & 1/30 & 10^{-5} & 4.61 & 1.14 & 10.6 & 65.7 & 39.1  \cr
          1/4 & 1/12 & 10^{-30} & 9.73 & 1.16 & 47.4 & 139 & 103 \cr
          1/4 & 1/12 & 10^{-5} & 4.40 & 0.468 & 9.64 & 25.9 & 8.01 \cr
 }}
The chosen values of $a_i$ correspond to pure $G_2 \times E_8$
resp.\ $G_2 \times E_6$ gauge theory. Again, the critical point is
a minimum of the potential.

The fact that $a_1 > a_2$ is needed to have a solution (lying well
inside the K\"ahler cone) can be seen directly from \volswp: the
approximate solution is $\tau_i = - \ln W_0 / 2 \pi a_i$, but
\volswp\ implies $\tau_1<\tau_2$, so $a_1 > a_2$.

\newsec{Conclusions}

Our primary result is that we have candidate
\IIb\ orientifold compactifications
in which nonperturbative effects will stabilize all
complex structure, K\"ahler and dilaton moduli.

Models which stabilize all K\"ahler moduli by D$3$-instanton effects
are not generic, but not uncommon either.  We listed all the
possibilities with Fano threefold base, and many possibilities whose
bases are $\IP^1$ fibrations.  It turns out that these models must
have several K\"ahler moduli and several non-perturbative
contributions to the superpotential, as in the early racetrack
scenarios for moduli stabilization.

We see no obstacle to adding supersymmetry breaking effects such
as the antibrane suggested by KKLT, D term effects or others.
One also expects these potentials to have many F breaking minima
(statistical arguments for this are given in \dd).
On general grounds, since the configuration spaces parameterized
by other moduli such as brane and bundle moduli are compact,
after supersymmetry breaking all moduli should be stabilized.

The models with Fano threefold base are rather complicated, with
many complex structure moduli.  It is not clear to us that this
makes them less likely candidates to describe real world physics,
but it is certainly a problem when using them as illustrative examples.

There may well be simpler models among the $\IP^1$ fibered models,
whose nonperturbative effects have a simple gauge theory picture.
Perhaps the simplest is the $\IP^1$ fibration over $\IP^2$ or ``11169
model.''  While the standard F theory analysis of the models we
discussed have suggests that they have too much massless matter in one
gauge factor to produce non-perturbative superpotentials, we suspect
that other effects, in particular flux couplings to these brane
world-volumes, would lift this matter, leading to working models,
and intend to return to this in future work.

Another possibility for finding simpler models would be to
stabilize some of the K\"ahler moduli with D terms.  Following
lines discussed in \aspdoug, one can show that in configurations
containing branes wrapping $k$ distinct cycles, D terms will generally
stabilize $k-1$ relative size moduli.  Since this relies on $\alpha'$
corrections, the resulting values of K\"ahler moduli will be string
scale, but order one factors in the volumes and gauge field strengths
might be arranged to make this a few times the string scale, which
could suffice.

In any case, even the simplest models under discussion have many more
moduli to stabilize.  Furthermore, one must check that a candidate
vacuum is not just a solution, but has no tachyonic instability.  Now
once one has established that these moduli indeed parameterize a
compact configuration space, it is clear that the minimum of the
effective potential on this space will be a stable vacuum.  While it
may be hard to compute the value of the potential at this minimum, the
large number of flux vacua strongly suggests (as in \boupol) that
whatever it is, it can be offset by a flux contribution to lead to a
metastable de Sitter vacuum.  Thus, granting the effective potential
framework, it will be extremely surprising if vacua of this type do
not exist, while technically quite difficult to find them.  This is
the type of picture which motivates the statistical approach discussed
in \refs{\stat,\ad,\BanksES,\dd,\ddtwo}, as well as anthropic
considerations such as \refs{\barrow,\susskind,\BanksES}.

Can we say anything about the validity of effective theory?  At
present we see no clear reason from string/M theory or quantum gravity
to doubt it.  However, even within the effective field theory
framework, there is another important assumption in KKLT, our work,
and the other works along these lines.  Namely, we have done
Kaluza-Klein reduction in deriving the configuration space of
Calabi-Yau metrics, and typically will take similar steps throughout
the derivation of the effective field theory.  Could it be that in
many of these backgrounds, some of the KK and stringy modes which are
dropped in this analysis are in fact tachyonic?  Since these
constructions rely on approximate cancellations between many diverse
contributions to the vacuum energy, it is conceivable that subsectors
of the theory have instabilities which do not show up in the final
effective lagrangian; this should be examined.  In any case, a lot of
work remains to see whether these models are as plentiful as they now
appear.

\bigskip

We thank V. Balasubramanian, T. Banks, P. Berglund, V. Braun,
D.-E. Diaconescu, M. Dine, A. Grassi, S. Kachru, J. K\"appeli, G.
Moore, S. Sethi and S. Trivedi for valuable discussions and
correspondence. We are especially grateful to B. Acharya and R.
Reinbacher for their collaboration in the early stages of the
project.

This research was supported in part by DOE grant DE-FG02-96ER40959.
M.R.D. is supported by the Gordon Moore Distinguished Scholar
program at Caltech.

\appendix{A}{Counting flux vacua when $b_3$ is large}

Equation \Nflux\ gives the expected number of flux vacua with $L
\leq L_*$, $|W_0|^2 \leq \lambda_*$, $\tau \in \hat{\cal H}
\subseteq {\cal H}$ and $z \in \hat{\cal M}_c \subseteq {\cal
M}_c$, for $\lambda_* \ll L_*$. Here we will derive expressions
for $\rho_0$ suitable for Monte Carlo estimates in the case of
many moduli.

It is somewhat more convenient to rewrite \Nflux\ first as
 \eqn\Nfluxtwo{
 N_{vac} = {\rm vol}(\hat{\cal H}) \int_{\hat{\cal M}_c}
 d^{2n} \! z \, \det g \,\, \nu(z).
 }
The density function $\nu(z)$ is \dd:
 \eqn\nudef{
 \nu(z) = {2^{2n+2} \over \det g} \int_{|X|^2 \leq \lambda_*}
 d^2 \! X \int_{Z_i Z^i \leq L_*}
 d^{2n}\! Z \,
 \, |\det \left( \matrix{
  0 & Z_j \cr
  Z_i & {\cal F}_{ijk} {\bar{Z}}^k \cr } \right)|^2 + {\cal O}(X).
 }
The integration variables $X$ and $Z^i$ range over $\IC$ and are
the remnants of the flux vector $N$ after a certain $z$-dependent
change of basis and after imposing the constraint $DW_0=0$ (now
considered as a constraint on the continuous flux components at a
given point $z$). Indices are lowered with the metric $g_{i
\bar{j}}$. The factor in front of the integral comes from the
Jacobian corresponding to the change of basis, and the determinant
in the integrand is the $|\det D^2 W_0|$ Jacobian accompanying the
delta function imposing $DW_0=0$. Finally, the ${\cal F}_{ijk}$
are the ``Yukawa couplings'' characterizing the special geometry
of the complex structure moduli space, i.e.\
 \eqn\Fijkdef{
 {\cal F}_{ijk} = e^{{\cal K}_c} \int_Z \Omega \wedge \partial_i
 \partial_j \partial_k \Omega.
 }
Since we are interested in very small values of the cosmological
constant, i.e.\ $\lambda_* \ll 1$, the ${\cal O}(X)$ part can be
dropped from the integrand and the integral over $X$ simply gives
a factor $\pi \lambda_*$.

By rewriting the determinant in the integrand as a Gaussian
fermionic integral and the bosonic integral as a Laplace
transformed Gaussian \dd, and then doing the bosonic integral,
this expression can be rewritten as a fermionic integral with
quartic fermionic action, hence it reduces to a finite number of
terms. (Alternatively, one could stick to bosonic variables and
apply Wick's theorem.) This is useful for low $n$, but for large
$n$ the number of terms becomes enormous, about $n^{4n}$, and
straightforward numerical evaluation becomes impossible. Instead,
we want to rewrite \nudef\ in a form suitable for Monte Carlo
estimates.

To this end, we implement the constraint $\|Z\|^2 \leq L_*$ by
inserting $\int_0^{L_*} d\ell \, \delta(\ell - \|Z\|^2)$ in the
integral. Changing variables from $Z$ to $U$ with $Z^i =
\sqrt{\ell} \, U^i$, and doing the $\ell$-integral then gives
 \eqn\Nvacnu{
 \nu(z) = {2^{2n+2} \pi \lambda_* L_*^{2n+1} \over (2n+1) \det g} \int d^{2n} U
 \, \delta(\|U\|^2-1) \, |\det \left( \matrix{
  0 & U_j \cr
  U_i & {\cal F}_{ijk} {\bar{U}}^k \cr } \right)|^2.
 }
Now define the following ``spherical'' average, for any function
$f(U)$:
 \eqn\averagedef{
 \left\langle f(U) \right\rangle_{\|U\|=1} = {\int d^{2n} U
 \, \delta(\|U\|^2-1) \, f(U) \over
 \int d^{2n} U \, \delta(\|U\|^2-1).
 }
 }
Noting that
 \eqn\area{
 \int d^{2n} U \, \delta(\|U\|^2-1) = {\pi^n \over (n-1)! \det g}
 }
we can thus write
 \eqn\Nvactwo{
 \nu(z) = {2^{2n+2} \pi^{n+1} \lambda_* L_*^{2n+1} \over (2n+1)(n-1)!} \, \mu(z)
 }
with
 \eqn\mudef{
 \mu(z) = {1 \over (\det g)^2} \left\langle
 |\det \left( \matrix{
  0 & U_j \cr
  U_i & {\cal F}_{ijk} {\bar{U}}^k \cr } \right)|^2 \right\rangle_{\|U\|=1}.
 }
Comparing this to \Nflux, we get
 \eqn\rhoest{
 \rho_0(z) = {(2n+2) (2n)! \over \pi^{n+1}(n-1)!}\, \mu(z)
 }

This can be made more explicit in the large complex structure
limit of $Z$,
parametrized with the special coordinates $t^i=x^i+i \, y^i$, where
$y^i$ becomes large and $x_i \in [0,1]$. All data is encoded in
the triple intersection numbers $\tilde{D}_{ijk}$ of the mirror
$\tilde{Z}$ to $Z$. The K\"ahler potential is ${\cal K}_c = -
\ln(2^3 \tilde{V})$, where $\tilde{V}={1 \over 6} \tilde{D}_{ijk}
y^i y^j y^k$. The metric and its determinant are
 \eqn\LCSmetric{
 g_{i\bar{j}} = {\tV_i \tV_j \over 4 \tV^2} - {\tV_{ij} \over 4 \tV},
 \qquad
 \det g = {(-1)^{n+1} \det \tV_{ij} \over 2^{2n+1} \tV^n},
 }
where $\tV_i = \partial_i \tV = {1 \over 2} \tD_{ijk} y^j y^k$ and
$\tV_{ij} = \partial_i \partial_j \tV = \tD_{ijk} y^k$. Finally,
the Yukawa couplings are ${\cal F}_{ijk} = e^{{\cal K}_c}
\tD_{ijk}$. Using all this and pulling the factor $e^{{\cal K}_c}$
out of the determinant, \mudef\ becomes
 \eqn\mulcs{
 \mu(y) = {1 \over 4^{n-4}}{\tV^2 \over (\det \tV_{ij})^2}
 \left\langle
 |\det\left(
 \matrix{
 0 & U_j \cr
 U_i & \tD_{ijk} \bar{U}^k
 }
 \right)|^2
 \right\rangle_{\|U\|=1}
 }
Note that $\mu(y)$ is invariant under rescaling $y^i \to \lambda
y^i$.

The average $\langle f(U) \rangle$ can be estimated numerically
using Monte Carlo methods; for many variables this is in fact the
only possible way. In the simplest version, one repeatedly picks a
random vector $Z$ from a normal distribution with mean zero and
covariance matrix proportional to the metric $g_{i\bar{j}}$ (or
any other distribution depending only on $\|Z\|$), one evaluates
$f(U)$ with $U \equiv Z/\|Z\|$, and at the end one computes the
average of the values obtained. This gives an approximate value
for $\rho_0(y)$, the approximation becoming better with increasing
number of sampling points.

What remains to be computed then to get the number of flux vacua
is the integral \Nflux. Again this can be done by Monte Carlo
integration. However, computing $\rho_0(y)$ at every sample point
in the way described above would require far too much computation
time in the models we consider. Therefore we will replace
$\rho_0(y)$ by its average over a limited number of sample points.
This is a reasonable thing to do, since in examples $\rho_0(y)$
stays of more or less the same order of magnitude over the large
complex structure part of moduli space (in particular, as noted
before, it is scale invariant). The integral then becomes
proportional to the volume of the part of moduli space under
consideration, which can estimated numerically in reasonable time.

One important universal property of this integral can be deduced
directly: vacua are strongly suppressed at large complex
structure. To see this, note that under rescaling $y \to \lambda
y$, $d^n y \det g \to \lambda^{-n} d^n y \det g$, so if for
instance the region under consideration is given by $\tV>V_*$,
then
 \eqn\Rdep{
 N_{vac}(\tV>V_*) \propto 1/V_*^{n/3}.
 }
We discuss this in more detail in \ddtwo.

\appendix{B}{Taming the complex structure moduli space of the ${\cal F}_{18}$
model}

The Calabi-Yau 3-fold $Z$ of the ${\cal F}_{18}$ model of section
5.1 has 89 complex structure moduli. Describing this space
completely including its exact periods would be extremely complex,
so we restrict ourselves to the large complex structure
limit, which can be constructed as the classical K\"ahler moduli
space of the mirror $\tilde{Z}$. Even this poses quite a
challenge.

The classical K\"ahler moduli space of $\tilde{Z}$ is specified
entirely in terms of the triple intersection numbers of the
divisors $\tD_i$. These in turn can be computed from the quadruple
intersection numbers of the divisors of the ambient toric variety
$\tilde{Y}$. There are more than
100 independent divisors in $\tilde{Y}$, so a priori there are
more than $100^4 = 10^8$ intersection numbers to compute.

Before one can start doing this, one needs
a maximal triangulation of the fan of the toric variety. Although
the fan in this case is very large (116 points), we managed to do this by hand,
and found 576 cones of volume 1.

The next step is to determine the actual K\"ahler cone within the
$\IR^{89}$ parameter space, i.e.\ the cone in which all
holomorphic curves have positive area. This involves constructing
and solving 377 inequalities in 89 variables, which is more
difficult than one might expect. Even finding just one point
satisfying these inequalities takes several hours on a PC.

In the end we want to compute numerically the volume of a subspace of the
K\"ahler cone, as described in appendix A. Because of the high
dimensionality of the space, this needs to be done by Monte Carlo.
The integrand is relatively costly to evaluate, as it involves
computing the $89 \times 89$ determinant given in \LCSmetric. On a
2.4 GHz Pentium, evaluation using $10^7$ Monte Carlo sample
points took about three days.

In the following we outline how we proceeded to achieve these goals.

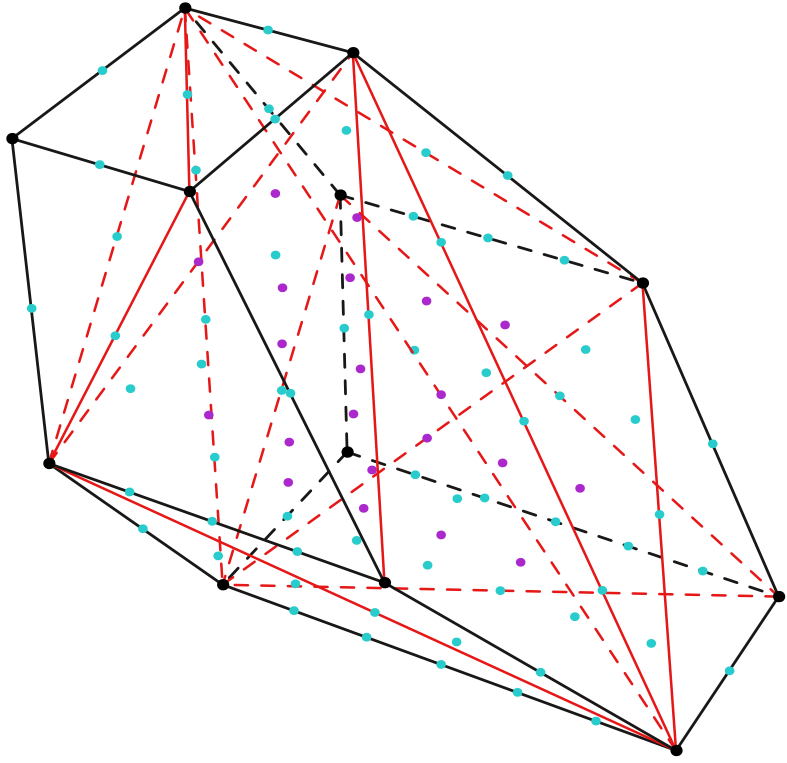
\ifig\sing{Partial triangulation of a 3-face of the polyhedron.
Black dots correspond to vertices, blue dots correspond to points
lying on edges or codimension two faces, while purple dots
correspond to points interior to the 3-face.
}{\epsfxsize3.8in\epsfbox{three.eps}}

\subsec{Intersections}

The toric variety $\tilde{Y}$ has 116 toric divisors $\tD_i$,
which correspond to points $p^i \in \IR^4$ in the polyhedron
$\nabla_{\tilde{Y}}$. The vertices of $\nabla_{\tilde Y}$ are
given by
 $$
 \eqalign{&{\tt
(4,0,-2,1),~~(-2,-2,0,1),~~(-2,2,0,1),~~(0,0,0,-1),~~(2,2,0,1),}\cr
&{\tt
(-2,0,2,1),~~(0,0,2,1),~~(-2,2,-2,1),~~(4,2,-2,1),~~(2,-2,0,1),}\cr
 &{\tt (-2,-2,2,1),~~(0,-2,2,1),~~(-2,0,-2,1).}}
 $$
The triangulation of the polytope has 576 cones $\tD_i \tD_j \tD_k \tD_l$,
all of volume 1.


The rule for the intersection product of four {\it distinct}
divisors is the following: if the four divisors span a cone, their
intersection is 1, otherwise it is 0. It is more complicated to
find intersections where some divisors are the same, such as
$\tD_i \cdot \tD_i \cdot \tD_j \cdot \tD_k$. This is done by
making use of the 4 linear equivalence relations that exist
between the 116 toric divisors, which can be simply read off from
the points: $p^i_\mu \tD_i = 0$. This gives four equations
$p^l_\mu \tD_{lijk} = 0$ for each distinct triple $ijk$. The
unknowns are the double index intersection numbers $\tD_{iijk}$,
so there are (more than) enough equations to find these. Solve this
using a computer as one system of millions of equations in
millions of variables would take a lot of time and memory. However, the
problem can be split up in a much smaller number of systems of
four equations in three variables: for a given $i<j<k$, the three
variables are $\tD_{iijk},\tD_{ijjk},\tD_{ijkk}$, and the only
$ijk$ which need to be taken into consideration are those for
which $\tD_i \tD_j \tD_k$ actually appears somewhere as a face in
the list of cones. For other $ijk$ the three unknowns are
trivially zero. A similar reasoning can be followed to compute
intersections with three and four identical indices. Thus almost
all intersections vanish, and the remaining ones can be computed
by computer in less than a minute.

Let us now turn to the Calabi-Yau hypersurface $\tilde{Z}$ in
$\tilde{Y}$. Of the 116 divisors $\tD_i$ in $\tilde{Y}$, only 93
intersect $\tilde{Z}$ and descend to divisors on the Calabi-Yau
(89 of those are independent). We will denote these divisors by
the same $\tD_i$. Their triple intersection numbers are now easy
to compute. They are simply given by the intersection with the
anticanonical divisor: $\tD_{ijk}=\sum_l \tD_{ijkl}$. The
resulting volume function is $\tV = {1 \over 6} \tD_{ijk} x^i x^j
x^k$.

Finally, the large complex structure prepotential of the original
Calabi-Yau $Z$ is given by this expression with the $x^i$
replaced by complex variables, and takes the form
$$
{\cal F}=\,{{t_1}}^3 - \,{{t_2}}^2\,{t_3} + 780\ {\rm more\ terms}.
$$
It is available upon request.

\subsec{Mori cone}

A class in $H_2(\tilde{Z})$ is specified by its intersections with
the divisors $\tD_i$. It is important to know which of these
classes can effectively be realized as holomorphic curves. The set
of such effective curves forms a cone, called the Mori cone. The
generators of the Mori cone $C_a$, given by their intersection
numbers $C_{ai}$ with the divisors $\tD_i$, are contained in the
list of `special' linear relations $C_{ai} p^i = 0$ between the
polyhedron points $p^i$ corresponding to the divisors. There is
one such special relation for each adjacent pair of cones in the
triangulation of the polyhedron, found as follows. Denote the
three common points of the two cones by $f_1$, $f_2$ and $f_3$,
and the two additional points by $p$ and $q$. Then there will be a
relation $p + q + n_1 f_1 + n_2 f_2 + n_3 f_3 = 0$, with the $n_i$
integer.

Applying this to our model gives 579 relations. These correspond
to effective curves of $\tilde{Y}$. What we want however are
effective curves of $\tilde{Z}$, and these are obtained by keeping
only the curves having zero intersections with the 23 divisors of
$\tilde{Y}$ which do not descend to divisors of $\tilde{Z}$. This
reduces the list to 377. Only a subset of this list will
constitute a basis of generators of the Mori cone: any curve that
is a positive linear combination of the others can be dropped. A
basis can thus be constructed in steps as follows. Start with the
first curve. Add to this the second one and check if in the
resulting set any one of the curves is a positive multiple of the
other. If so, remove this curve. Then add the third curve and
remove any curve in the resulting set that is a positive linear
combination of the others. Then add the fourth curve, and so on,
till all 377 candidates have been considered. This procedure is
much faster than starting with all curves and removing the
dependent ones one by one, because verifying if a given vector
equals some {\it positive} linear combination of a set of $n$
vectors becomes nontrivial if $n$ is bigger than the dimension,
and with $n$ substantially bigger, it is computationally extremely
expensive (hours for $n=377$ in this case).

The resulting basis of the Mori cone consists of 111 generators
$C_a$.

\subsec{K\"ahler cone}

The dual of the Mori cone is the K\"ahler cone, that is the set of
all $J = x^i D_i$ such that $C_a \cdot J>0$. Note that this is a
tiny fraction of the $\IR^{89}$ parameter space; an estimate for
the probability of a random point to be in the K\"ahler cone is
$(1/2)^{111} \sim 10^{-34}$. Since we want to integrate over the
K\"ahler cone, it is important to have a good parametrization of
it --- multiplying the integrand by a step function with support
on the K\"ahler cone and Monte Carlo integrating over all $x$
certainly won't do the job, since effectively all sample points
will integrate to zero.
Ideally, one would construct the exact generators $K_p$ of the
K\"ahler cone and write $J = t^p K_p$, $t^p>0$. There exists an
algorithm to find these generators, decribed in \fulton\ p.11.
Unfortunately, because the Mori cone is far from simplicial, this
involves running over ${111 \choose 88} \sim 10^{23}$ candidate
generators (the rays obtained by intersecting 88 of the 111 zero
planes). It would take about the age of the universe to complete
this task on a PC. Nevertheless, with some luck, we were able to
construct an approximate parametrization. Applying Mathematica's
function {\tt InequalityInstance} to find one solution to our set
of 111 inequalities results in a point which has exactly 89 curve
areas $C_i \cdot J$ equal to 1, and 22 bigger than 1. Taking the
areas of these special curves as coordinates $y_i$, it turns out
that for uniformly random positive $y$ values, the resulting $J$
generically lies inside the Kahler cone. The integral over the
K\"ahler cone can therefore be done by Monte Carlo integrating
over all positive $y$ and multiplying the integrand by a step
function with support on the points satisfying the remaining 22
inequalities. In this way, not too many sample points evaluate to
zero.

\listrefs
\end